\documentclass[acmsmall,screen]{acmart}
\setcopyright{rightsretained}
\acmPrice{}
\acmDOI{10.1145/3498663}
\acmYear{2022}
\copyrightyear{2022}
\acmSubmissionID{popl22main-p9-p}
\acmJournal{PACMPL}
\acmVolume{6}
\acmNumber{POPL}
\acmArticle{2}
\acmMonth{1}

\usepackage{booktabs}   
\usepackage{subcaption} 
\usepackage{amsmath}
\usepackage{qcircuit}
\usepackage{stmaryrd}
\usepackage{mathrsfs}
\usepackage{enumerate}
\usepackage{tikz-cd}
\usepackage{tikzit}

\tikzstyle{scalar}=[fill=white, draw=black, shape=circle, tikzit fill=white, tikzit draw=black]
\tikzstyle{map}=[fill=white, draw=black, shape=rectangle, tikzit fill=white, tikzit draw=black, minimum width=4em]
\tikzstyle{bullet}=[fill=black, draw=none, shape=circle, tikzit fill=black, inner sep=1pt]
\tikzstyle{map1}=[fill=white, draw=black, shape=rectangle, tikzit fill=white, tikzit draw=black, minimum width=2em]
\tikzstyle{state}=[fill=white, draw=black, tikzit fill=white, tikzit draw=black, regular polygon, regular polygon sides=3, shape border rotate=180]
\tikzstyle{map3}=[fill=white, draw=black, shape=rectangle, tikzit fill=white, tikzit draw=black, minimum width=6em]
\tikzstyle{circle}=[fill=white, draw=black, shape=circle, tikzit fill=white, tikzit draw=black, inner sep=1pt]
\tikzstyle{label}=[font={\tiny}]
\tikzstyle{effect}=[fill=white, draw=black, shape=circle, tikzit fill=white, tikzit draw=black, regular polygon, regular polygon sides=3]
\tikzstyle{clone}=[fill=black, draw=black, shape=circle, inner sep=2pt]
\tikzstyle{map4}=[fill=white, draw=black, shape=rectangle, tikzit fill=white, tikzit draw=black, minimum width=9em]

\tikzstyle{invis}=[-, dashed, dash pattern=on 0.2mm off 0.2mm]

\usepackage{physics}
\usepackage{quiver}
\usepackage{color}

\newcommand{\tot}{\xrightarrow}
\newcommand{\sem}[1]{\ensuremath{\llbracket #1 \rrbracket}}

\newcommand{\opnorm}[1]{\norm{#1}_\mathrm{op}}

\newcommand{\cat}[1]{\ensuremath{\mathbf{#1}}}
\newcommand{\opp}{{\ensuremath{{}^{\mathrm{op}}}}}

\newcommand{\Unitary}{\cat{Unitary}}
\newcommand{\Isometry}{\cat{Isometry}}
\newcommand{\CPTP}{\cat{CPTP}}

\newcommand{\FinSet}{\cat{FinSet}}
\newcommand{\FinBij}{\cat{FinBij}}

\newcommand{\Qbit}{\ensuremath{\mathit{Qbit}}}
\newcommand{\Bit}{\ensuremath{\mathit{Bit}}}

\newcommand{\id}{\mathrm{id}}
\newcommand{\inl}{\mathrm{inl}}
\newcommand{\inr}{\mathrm{inr}}

\newcommand{\RR}[1]{\ensuremath{R[#1]}}
\newcommand{\LL}[1]{\ensuremath{L[#1]}}
\newcommand{\LR}[1]{\LL{\RR{#1}}}

\newcommand{\RC}{\RR{\cat{C}}}

\newcommand{\EE}{\ensuremath{\mathcal{E}}}
\newcommand{\DD}{\ensuremath{\mathcal{D}}}

\newcommand{\UPi}{\ensuremath{\mathcal{U}\Pi}}
\newcommand{\UPia}{\ensuremath{\mathcal{U}\Pi_a}}
\newcommand{\UPichia}{\ensuremath{\mathcal{U}\Pi^\chi_a}}

\def\fcmp{\mathbin{\raise 0.6ex\hbox{\oalign{\hfil$\scriptscriptstyle      \mathrm{o}$\hfil\cr\hfil$\scriptscriptstyle\mathrm{9}$\hfil}}}}

\newcommand{\seqq}{\fcmp}
\newcommand{\comb}{\mathit}
\newcommand{\fromto}{\leftrightarrow}
\newcommand{\of}{\mathbin{:}}
\newcommand{\inv}{\comb{inv}}
\newcommand{\ato}{\rightarrowtail}
\newcommand{\acto}{\rightsquigarrow}
\newcommand{\acmp}{\mathbin{>\!\!>\!\!>}}
\newcommand{\ppp}{+\!\!+\!\!+}
\newcommand{\ttt}{\ast\!\!\ast\!\!\ast}

\newcommand{\trans}[1]{\ensuremath{\mathcal{T}\sem{#1}}}

\newtheorem{theorem}{Theorem}
\newtheorem{lemma}[theorem]{Lemma}
\newtheorem{proposition}[theorem]{Proposition}

\theoremstyle{definition}
\newtheorem{definition}[theorem]{Definition}

\definecolor{silver}{rgb}{0.9,0.9,0.9}
\definecolor{palegreen}{rgb}{0.8,0.95,0.8}
\definecolor{paleyellow}{rgb}{0.95,0.95,0.8}

\newif\ifarxiv

\begin{document}

\title{Quantum Information Effects}

\author{Chris Heunen}
\affiliation{
  \department{School of Informatics}
  \institution{University of Edinburgh}
  \streetaddress{10 Crichton Street}
  \city{Edinburgh}
  \postcode{EH8 9AB}
  \country{United Kingdom}
}
\email{chris.heunen@ed.ac.uk}

\author{Robin Kaarsgaard}
\affiliation{
  \department{School of Informatics}
  \institution{University of Edinburgh}
  \streetaddress{10 Crichton Street}
  \city{Edinburgh}
  \postcode{EH8 9AB}
  \country{United Kingdom}
}
\email{robin.kaarsgaard@ed.ac.uk}

\begin{abstract}
  We study the two dual quantum information effects to manipulate the amount of information in quantum computation: hiding and allocation. The resulting type-and-effect system is fully expressive for irreversible quantum computing, including measurement.
  We provide universal categorical constructions that semantically interpret this arrow metalanguage with choice, starting with any rig groupoid interpreting the reversible base language.
  Several properties of quantum measurement follow in general, and we translate  (noniterative) quantum flow charts into our language.
  The semantic constructions turn the category of unitaries between Hilbert spaces into the category of completely positive trace-preserving maps, and they turn the category of bijections between finite sets into the category of functions with chosen garbage.
  Thus they capture the fundamental theorems of classical and quantum reversible computing of Toffoli and Stinespring.
\end{abstract}

\ccsdesc[500]{Theory of computation~Quantum computation theory}
\ccsdesc[500]{Theory of computation~Categorical semantics}

\keywords{quantum computation, reversible computation, information effects, measurement, effects, arrows, categorical semantics}

\maketitle

\begin{acks}
  This material is based upon work supported by the
  \grantsponsor{EPSRC}{Engineering and Physical Sciences Research
  Council}{https://epsrc.ukri.org} Fellowship
  No.~\grantnum{EPSRC}{EP/R044759/1}, and the \grantsponsor{DFF}{Independent
  Research Fund Denmark}{https://dff.dk} under DFF-International Postdoc
  Fellowship No.~\grantnum{DFF}{0131-00025B}.
  We thank Pablo Andrés-Martínez and the anonymous reviewers for their comments,
  corrections, and suggestions for this paper, and Martti Karvonen for early 
  discussions.
\end{acks}

\section{Introduction}
Something is rotten in the state of quantum computing. It subsumes classical computing, which is generally irreversible, yet it is most often formulated as a reversible quantum circuit, with an irreversible quantum measurement as an afterthought. The conceptual status of this irreversible measurement remains mysterious. This is known as the measurement problem.

Classical computing itself is most often formulated as composed of irreversible operations. However, by the seminal works of Toffoli~\cite{toffoli:reversible} and Bennett~\cite{bennett:logicalrev}, and more recently by James and Sabry~\cite{jamessabry:infeff}, we know that it can also be phrased in terms of reversible operations, as long as we consider systems to be open and interact with an environment that is eventually disregarded. This final part is important, as reversible computations alone (be they classical or quantum) cannot change the \emph{amount} of information (as measured by an appropriate notion of \emph{entropy}).
To understand the nature of quantum measurement in computation requires us to close two conceptual gaps:
\begin{enumerate}[(i)]
  \item from reversible classical computing to reversible quantum computing; and
  \item from reversible quantum computing to irreversible quantum computing.  
\end{enumerate}
It may seem that much has to be added to a reversible language to make it suitable for quantum computing. Similarly, it may seem that much less can be expressed in purely reversible quantum computations than in arbitrary quantum computations with measurements. We argue, however, that both gaps are smaller than they may appear.

To do this, we start with the reversible combinator language $\Pi$, which  governs classical reversible computation, and extend it with combinators for quantum phases and the quantum \emph{Hadamard gate}. We call the result $\UPi$ (``yuppie''), because it is already approximately universal for reversible quantum computing with unitary gates. 

To address (ii) we introduce two \emph{quantum information effects} --  computational effects that manipulate the amount of information -- through two arrows~\cite{hughes:arrows}. The first computational effect allows \emph{allocation} of auxiliary space on a hidden heap, leading to the arrow metalanguage $\UPia$ (``yuppie-a''). This calculus is approximately universal for quantum computing with \emph{isometries} rather than unitaries.
The second computational effect dually allows \emph{hiding}, by disregarding specifically marked \emph{garbage output}, leading to $\UPichia$ (``yuppie-chi-a''). We prove that this calculus is approximately universal for \emph{arbitrary quantum computations}, including  measurement.

Thus we have an arrow metalanguage that, with two simple computational effects on top of a pure reversible model, is fully expressive for irreversible quantum computing. All the allocation and hiding is tracked by the type system, and so allows us to compile an irreversible quantum program into an explicit reversible quantum circuit.

We provide matching categorical semantics via surprisingly simple concrete constructions that have very general universal properties. Vanilla $\Pi$ may be interpreted in rig categories: categories with two monoidal structures $(\otimes, \oplus)$, where the product ($\otimes$) distributes over the sum ($\oplus$). We will interpret $\UPi$ in the category $\cat{Unitary}$ of Hilbert spaces and unitaries. This is a choice of canonical model: all that is needed is a rig category with morphisms to interpret phase gates and the Hadamard gate (which we will see is equivalent to having a notion of \emph{superposition}).

For $\UPia$ we provide a free construction that turns a rig category $\cat{C}$ into a new one $\RR{\cat{C}}$ where the unit for the sum becomes initial. Then $\RR{\cat{Unitary}}$ is the category $\cat{Isometry}$ of Hilbert spaces and isometries. Dually, we interpret $\UPichia$ via a free construction $L$ making a monoidal unit terminal. Now $\LL{\cat{Isometry}}$ is the category $\cat{CPTP}$ of arbitrary (irreversible) quantum channels. Classically, $R$ transforms the category of bijections between sets into that of injections, and in turn $L$ transforms that into arbitrary functions with chosen garbage. This lets us reformulate Toffoli's \emph{fundamental theorem of reversible computing}~\cite{toffoli:reversible} as a purely categorical statement.

Surprising mileage is obtained from these simple constructions, as we prove in general several properties of measurement that can be expressed entirely as semantic equivalences between program fragments. For example, we show that measurement commutes with injections and projections, and that measurement is idempotent. More generally, we show that our setting can interpret (noniterative) \emph{quantum flow charts}~\cite{selinger:qpl}. This highlights the potential of $\UPichia$ not only as a vehicle for theoretical studies, here the study of measurement, but also as an intermediate language with a strong equational theory owing to its categorical semantics.
All constructions and translations in this paper are formalised in (heavily extended) Glasgow Haskell.

\subsubsection*{Related work}

Classical information effects are due to~\cite{jamessabry:infeff}, while the characterisation of $\Isometry$ and $\CPTP$ as successive completions of $\Unitary$ was first given by \cite{huotstaton:universal,huotstaton:completion} (see also \cite{heunenkaarsgaard:bennettstinespring}).
Quantum programming languages are an active research topic~\cite{bichselbaadergehrvechev:silq, selinger:qpl, paykinrandzdancewic:qwire, greenlumsdainerossselingervaliron:quipper,sabryvalironvizzotto:quantumctrl}.
In particular, quantum measurement has been studied extensively as a computational effect~\cite{vizzotto:quantumarrows1, vizzotto:quantumarrows2, altenkirch:quantumio, green:reversible, westerbaan:kleisli} too. While proven practically useful, the precise meaning of \emph{measurement-as-an-effect} has remained unclear, perhaps partly because of the wide-spread conscription to the view~``quantum data, classical control''~\cite{selinger:qpl}.

We generalise this previous work by providing the missing origin story of measurement-as-an-effect as a sequence of arrow constructions (semantically, categorical completions) that can be applied (and given precise meaning) to \emph{any} rig groupoid. In other words, where previous work shows \emph{that} measurement arises in this way, using facts specific to quantum theory, we show \emph{how}, using only the language of rig categories. With measurement thus defined in the abstract,
our categorical constructions eliminate the need for involved functional-analytic semantics using operator algebras~\cite{chowesterbaan:vonneumann, rennelastaton:enriched, pechouxperdrixrennelazamdzhiev:causality}.

The $\RR{-}$-construction on small categories was studied as a special case of freely adjoining monoidal indeterminates in \cite{hermidatennent:indeterminates}. An unquotiented, bicategorical version of this construction was also given here, later rediscovered in \cite{fongspivaktuyeras:backprop,cruttwellgavranovicghaniwilsonzanasi:gradient} as the $\mathrm{Para}$-construction. Though quotienting will be important here, a bicategorical construction could see use to study the structure and order of dilations (see also \cite{houghtonlarsen:dilations}).

\subsubsection*{Overview}

Section~\ref{sec:background} recalls background material. 
Next, Section~\ref{sec:yuppies} discusses $\Pi$, introduces the languages $\UPi$, $\UPia$, and $\UPichia$, and proves expressivity theorems.
Section~\ref{sec:semantics} deals with categorical semantics: it recalls rig groupoids, introduces the $L$ and $R$ constructions, and proves that they respect $\otimes$ and give the appropriate setting to interpret $\oplus$ as an arrow with choice. 
Section~\ref{sec:properties} concerns universal properties of the $L$ and $R$ constructions, and shows that they encompass two fundamental results: Toffoli's fundamental theorem, and Stinespring's dilation theorem.
In Section~\ref{sec:applications} we derive extra properties in the arrow metalanguage, generically in $\LL{\RR{\cat{C}}}$.
Finally, Section~\ref{sec:conclusion} concludes and lists interesting directions for future work.

\section{Background}\label{sec:background}

This section recalls the basics of quantum theory, monoidal categories, and information effects.

\subsection{Quantum theory}

For more details we refer to~\cite{yanofskymannucci:quantumcomputing,nielsenchuang:quantum}.

\subsubsection{Pure quantum theory and bra-ket notation} 
\label{ssub:pure_quantum_theory_and_bra_ket_notation}

A quantum system is captured by a complex Hilbert space $H$. For example, qubits are modelled by $H=\mathbb{C}^2$. The \emph{pure states} $\psi$ of $H$ are the vectors of unit norm: $\|\psi\|=1$. By convention, vectors are denoted as a \emph{ket} $\ket{\psi}$. This is handy, because then the functional $H \to \mathbb{C}$ that maps $\phi$ to the inner product $\innerproduct{\psi}{\phi}$ can be denoted as the \emph{bra} $\bra{\psi}$. As a further consequence, the outer product of $\phi$ with $\psi$ (with signature $H \to H$) becomes $\dyad{\phi}{\psi}$.

Pure dynamics of a quantum system are reversible. Evolution is given by a \emph{unitary} linear map $U \colon H \to H$, meaning that $U$ is a bijection that is \emph{isometric}: $\innerproduct{U\phi}{U\psi}=\innerproduct{\phi}{\psi}$.
More generally, any continuous linear function $f \colon H \to K$ has an \emph{adjoint} $f^\dag \colon K \to H$ satisfying $\innerproduct{f\phi}{\psi} = \innerproduct{\phi}{f^\dag \psi}$. An isometry then satisfies $f^\dag \circ f = \mathrm{id}$, and a unitary furthermore satisfies $f \circ f^\dag = \mathrm{id}$.

Pure quantum theory subsumes reversible classical computation. Any finite set $I$ generates a Hilbert space $\mathbb{C}^I$ of linear combinations of elements of $I$. Thus $\{ \ket{i} \mid i \in I \}$ forms a basis of $\mathbb{C}^I$ that is moreover orthonormal: $\innerproduct{i}{j}$ is 1 when $i=j$ and vanishes otherwise. We call this basis of $\mathbb{C}^n$ induced by the set $\{1,2,\ldots,n\}$ the \emph{computational basis}. Any bijection of $\{1,2,\ldots,n\}$ induces a unitary on $\mathbb{C}^n$ that preserves the computational basis.

If two quantum systems are modelled by Hilbert spaces $H$ and $K$, the compound system is given by their tensor product $H \otimes K$. For example, a 3-qubit system is modelled by $\mathbb{C}^2 \otimes \mathbb{C}^2 \otimes \mathbb{C}^2 \simeq \mathbb{C}^8$. Similarly, if $H$ and $K$ evolve along unitaries $U$ and $V$, then $H \otimes K$ evolves along $U \otimes V$.

\subsubsection{Mixed quantum theory} 
\label{ssub:pure_vs_mixed_quantum_theory}

A quantum computation (in the quantum circuit model) consists of a composition of tensor products of unitary gates, which is entirely reversible. However, reading out the result of the computation requires a measurement, which is an irreversible operation. The standard model therefore considers \emph{mixed states}. These are given by a \emph{density matrix}, which is a linear function $\rho \colon H \to H$ such that $0 \leq \innerproduct{\rho(\psi)}{\psi} \leq 1$ for all $\ket{\psi}$. Thus any pure state $\ket{\psi} \in H$ is also a mixed state $\dyad{\psi}{\psi} \colon H \to H$. 

Mixed states no longer have reversible dynamics. Any unitary $U \colon H \to H$ still induces a map that takes a mixed state $\rho$ to a mixed state $U^\dag \circ \rho \circ U$. But now the appropriate dynamics allow more possibilities, generally given by so-called \emph{completely positive trace-preserving (CPTP)} maps, also known as \emph{quantum channels}. It is not important here to set out their definition. What is important is \emph{Stinespring's dilation theorem}, which says that any CPTP map $H \to K$ may be factored as a pure evolution $H \to K \otimes G$, given by $\rho \mapsto V^\dag \circ \rho \circ V$ for an isometry $V$, followed by a map $K \otimes G \to K$. That is, irreversible (mixed) quantum theory is contained within reversible (pure) quantum theory, as long as you allow an environment to play the role of auxiliary state space but disregard it.

\subsubsection{Superposition and measurement}
Superposition is the ability of a quantum state $\ket\rho$ to occupy several classical states $\ket{b_1} \dots \ket{b_n}$ at once, so long as no measurement occurs. Each classical state in a superposed state is weighted by a complex number $\alpha_i$ known as an \emph{amplitude}. Once a system in superposition $\ket\psi = \sum_{i=1}^n \alpha_i \ket{b_i}$ is measured, it collapses to one of the classical states $\ket{b_k}$. The outcome of such a measurement is probabilistic, with the probability of observing $\ket{b_i}$ given by $|\bra{b_i}\ket{\psi}|^2$; this is called the \emph{Born rule}. Using density matrices, measurement with respect to  $\{\ket{b_i}\}_{i \in I}$ is represented by the \emph{measurement instrument channel}
\[
  \rho \mapsto \sum_{i \in I} \dyad{b_i}{b_i} \rho \dyad{b_i}{b_i}
\]
that sends quantum states to their post-measurement (mixed classical) states. For example, measuring a qubit $\dyad{\psi}{\psi}$ for $\ket{\psi} = \alpha \ket0 + \beta \ket1$ with respect to $\{\ket0, \ket1\}$ results in the mixed state $|\alpha|^2 \dyad{0}{0} + |\beta|^2 \dyad{1}{1}$.

\subsubsection{Global and relative phase}
Recall that the complex conjugate of a complex number $\varphi = a+bi$ is $\overline{\varphi} = a-bi$. A \emph{phase} is a complex number satisfying $\varphi \cdot \overline{\varphi} = 1$; equivalently, $\varphi$ has norm $1$. Quantum states that differ only by a \emph{global phase}, $\ket{\psi'} = \varphi \ket{\psi}$, are indistinguishable, in that they have the same measurement statistics. But the phase difference between \emph{parts} of  states can be incredibly important. For example, the states $\ket{+} = \frac{1}{\sqrt{2}}(\ket0 + \ket1)$ and $\ket{-} = \frac{1}{\sqrt{2}}(\ket0 - \ket1)$ differ only by the phase $-1$ in their amplitude for $\ket1$. This difference in \emph{relative phase} makes $\ket{+}$ and $\ket{-}$ orthogonal.

\subsection{Monoidal categories}

For our semantics, we will assume that the reader is familiar with the basic notions of categories and functors~\cite{leinster:categories}. To set notation, recall that a category $\cat{C}$ is symmetric monoidal when it comes equipped with a tensor product $\otimes \colon \cat{C} \times \cat{C} \to \cat{C}$, a unit object $I$, isomorphisms $\lambda_A \colon I \otimes A \to A$ and $\rho_A \colon A \otimes I \to A$ called unitors, isomorphisms $\alpha_{A,B,C} \colon A \otimes (B \otimes C) \to (A \otimes B) \otimes C$ called associators, and isomorphisms $\sigma_{A,B} \colon A \otimes B \to B \otimes A$ called symmetries for all objects $A,B,C$, that satisfy certain coherence laws~\cite{heunenvicary:cqt}.

A category lets one compose morphisms `in sequence'; a monoidal category additionally lets one compose morphisms `in parallel'. This is expressed satisfyingly in the graphical calculus for monoidal categories. We draw a morphism as a box with an incoming wire labelled by its domain and an outgoing wire labelled by its codomain. Composition becomes stacking boxes vertically, whereas we draw the tensor product of boxes side by side. In particular, objects $A \otimes B$ may be drawn as a single wire labelled $A \otimes B$, or as two parallel wires labelled by $A$ and $B$, and the nullary case of a wire labelled $I$ is simply not drawn. The special morphisms $\sigma_{A,B}$ are drawn as crossing two wires. The coherence laws simply say that one may ignore the coherence isomorphisms graphically. 
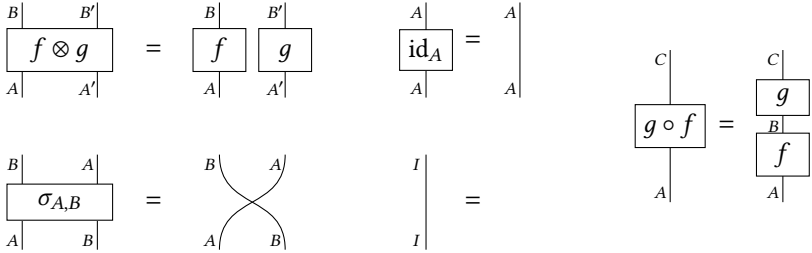
\ctikzfig{graphicalcalculus}
Of special interested are symmetric monoidal categories whose tensor unit is initial or terminal. If there is a unique morphism $A \to I$ for any object $A$, the monoidal category is called \emph{affine}, and if there is a unique morphism $I \to A$ for any object $A$, it is called \emph{coaffine}.

A functor $F \colon \cat{C} \to \cat{D}$ between symmetric monoidal categories is (strong) monoidal when it is equipped with isomorphisms $F(A) \otimes F(B) \simeq F(A \otimes B)$ and $F(I) \simeq I$ that respect the coherence isomorphisms of $\cat{C}$ and $\cat{D}$. It is \emph{strict} monoidal when these isomorphisms are in fact identities. Monoidal functors between (co)affine categories automatically preserve the terminal (initial) object.

\subsection{Information effects}

Classical computation, embodied by functions on finite sets, is irreversible, because applying a function in general loses information. This can be made precise via the Shannon entropy $H=-\sum p_i \log p_i$ that measures how surprising it is when a variable takes the value $i$ with probability $p_i$. The functions that preserve Shannon entropy are precisely bijections. This direct connection between information preservation and reversibility is a consequence of \emph{Landauer's principle}~\cite{landauer:irreversibility}.

The central idea of information effects~\cite{jamessabry:infeff} is that this irreversible model of computation arises from a reversible (bijective) model of computation, together with computational effects for duplicating and erasing information. Thus irreversible programs are reversible instructions governed by an arrow metalanguage that tracks interaction with a global environment.

Quantum theory, embodied by quantum channels between finite-dimensional Hilbert spaces, is also irreversible. The information content of a quantum state $\rho$ can be made precise by von Neumann entropy $S=-\mathrm{tr}(\rho \log \rho)$. In this case, the information-preserving maps are also the reversible ones: those of the form $\rho \mapsto U^\dag \circ \rho \circ U$ for unitary $U$. This mirrors the classical connection between information preservation and reversibility.

\section{Three generations of yuppie}\label{sec:yuppies}
$\Pi$ is a reversible combinator language introduced in \cite{bowmanjamessabry:dagger, jamessabry:infeff} to study strongly typed reversible \emph{classical} programming. Many extensions exist, such as partiality and iteration~\cite{bowmanjamessabry:dagger, jamessabry:infeff}, fractional types~\cite{chenchoudhurycarettesabry:fractional, chansabry:negativefractional}, negative types~\cite{chansabry:negativefractional}, and higher combinators~\cite{carettesabry:weakriggrpds, kaarsgaardveltri:engarde}. This section introduces a quantum extension to $\Pi$, and shows it to be approximately universal for unitaries, the canonical model of pure quantum computation (without measurement). We then use two arrow constructions to extend this with the quantum information effects of \emph{allocation} and \emph{hiding} to arrive at an arrow metalanguage which we prove approximately universal for quantum channels, the canonical model of full quantum computation (with measurement).

\subsection{Reversible classical combinators: $\Pi$}
\begin{figure}
  \fbox{
  \begin{minipage}{0.95\textwidth}
  \textbf{Syntax}
  \begin{align*}
    b & ::= 0 \mid 1 \mid b + b \mid b \times b & \text{(base types)} \\
    t & ::= b \leftrightarrow b & \text{(combinator types)} \\
    a & ::= \comb{id} \mid \comb{swap}^+ \mid \comb{unit}^+ \mid \comb{uniti}^+
     \mid \comb{assoc}^+  \mid \comb{associ}^+ \\
     & \enspace \mid \comb{swap}^\times \mid \comb{unit}^\times \mid 
     \comb{uniti}^\times
     \mid \comb{assoc}^\times \mid \comb{associ}^\times \\
     & \enspace \mid \comb{distrib} \mid \comb{distribi} \mid 
     \comb{distribo} \mid \comb{distriboi} & \text{(primitive combinators)} \\
    d & ::= \comb{midswap}^+ \mid \comb{midswap}^\times & \text{(derived 
    combinators)} \\
    c & ::= a \mid c \seqq c \mid c + c \mid c \times c & \text{(combinators)}
  \end{align*}
  \textbf{Typing rules}
  \begin{equation*}
    \begin{array}{c c c c c c}
      \comb{id} & \of & b \fromto b & \of & \comb{id} \\
      \comb{swap}^+ & \of & b_1 + b_2 \fromto b_2 + b_1 & \of & \comb{swap}^+ \\
      \comb{unit}^+ & \of & b + 0 \fromto b & \of & \comb{uniti}^+ \\
      \comb{assoc}^+ & \of & (b_1 + b_2) + b_3 \fromto b_1 + (b_2 + b_3) & 
      \of & \comb{associ}^+ \\
      \comb{swap}^\times & \of & b_1 \times b_2 \fromto b_2 \times b_1 & \of &
      \comb{swap}^\times \\
      \comb{unit}^\times & \of & b \times 1 \fromto b & \of 
      & \comb{uniti}^\times \\
      \comb{assoc}^\times & \of & (b_1 \times b_2) \times b_3 \fromto b_1 \times
      (b_2 \times b_3) & \of & \comb{associ}^\times \\
      \comb{distrib} & \of & 
      b_1 \times (b_2 + b_3) \fromto (b_1 \times b_2)
      + (b_1 \times b_3) & \of & \comb{distribi} \\
      \comb{distribo} & \of & b \times 0 \fromto 0 & \of & \comb{distriboi} \\
      \comb{midswap}^+ & \of & (b_1 + b_2)+(b_3 + b_4) \fromto 
      (b_1 + b_3)+(b_2 + b_4) & \of & \comb{midswap}^+ \\
      \comb{midswap}^\times & \of & (b_1 \times b_2) \times (b_3 \times b_4) 
      \fromto (b_1 \times b_3) \times (b_2 \times b_4) & \of & 
      \comb{midswap}^\times
    \end{array}
  \end{equation*}
  \begin{equation*}
    \frac{c_1 \of b_1 \fromto b_2 \quad c_2 \of b_2 \fromto b_3}{c_1 \seqq c_2 
    \of b_1 \fromto b_3} \qquad
    \frac{c_1 \of b_1 \fromto b_3 \quad c_2 \of b_2 \fromto b_4}{c_1 + c_2 
    \of b_1 + b_2 \fromto b_3 + b_4} \qquad
    \frac{c_1 \of b_1 \fromto b_3 \quad c_2 \of b_2 \fromto b_4}{c_1 \times c_2 
    \of b_1 \times b_2 \fromto b_3 \times b_4}
  \end{equation*}
  \end{minipage}
  }
  \caption{The syntax and type system of $\Pi$.}
  \label{fig:pi}
\end{figure}
The syntax and type system of the unextended calculus $\Pi$ is shown in
Fig.~\ref{fig:pi}. It comprises a small set of invertible, first-order,
strongly typed polymorphic combinators on data constructed from (classical) sum and products types, as well as their units $0$ and $1$. These combinators
enable data of sum and product type to be swapped (sending $\inl~x$ to $\inr~x$
and vice versa for sums, and $(x,y)$ to $(y,x)$ for products), reassociated,
and have their respective units added and removed in the usual way. Products
can also be distributed over sums (and back again) as usual. Finally, these
combinators can be composed in sequence $c_1 \seqq c_2$ and in parallel using
both $+$ and $\times$. That is, $c_1 \times c_2$ takes a pair $(x,y)$ and
produces the pair $(c_1~x, c_2~x)$, while $c_1 + c_2$ takes $\inl~x$ to
$\inl~(c_1 x)$ and $\inr~y$ to $\inr~(c_2 y)$.

Aside from the base combinators, a pair of useful derived combinators  $\comb{midswap}^+ \of (b_1 + b_2)+(b_3 + b_4) \fromto (b_1 + b_3)+(b_2 + b_4)$ and $\comb{midswap}^\times \of (b_1 \times b_2) \times (b_3 \times b_4) \fromto (b_1 \times b_3) \times (b_2 \times b_4)$ can be defined as 
\begin{align*}
  \comb{midswap}^+ & = \comb{assoc}^+ \seqq (\comb{id} + \comb{associ}^+) \seqq
  (\comb{id} + (\comb{swap}^+ + id)) \seqq (\comb{id} + \comb{assoc}^+) \seqq
  \comb{associ}^+ \\
  \comb{midswap}^\times & = \comb{assoc}^\times \seqq (\comb{id} \times
  \comb{associ}^\times) \seqq (\comb{id} \times (\comb{swap}^\times \times id))
  \seqq (\comb{id} \times \comb{assoc}^\times) \seqq \comb{associ}^\times
  \enspace.
\end{align*}
The definition and use of derived combinators, which we will make heavy use of, should be taken as no more than aliasing, or macro definition and expansion. Recursive systems (mutually or otherwise) of derived combinators are \emph{not} permitted.

$\Pi$ takes semantics in \emph{rig groupoids} (see Section~\ref{sec:semantics}), the canonical choice being the category $\cat{FinBij}$ of finite sets and bijective functions. Indeed, $\Pi$ is universal for finite bijective functions; Fig.~\ref{fig:gates} shows the implementations of the universal gate set $\{PX, CNOT, TOFFOLI\}$~\cite{toffoli:reversible}. 

\subsubsection{Inversion} 
\label{ssub:inversion}
\begin{figure}
\fbox{
\begin{minipage}{0.95\textwidth}
\begin{equation*}
  \begin{array}{r c l r c l}
    \inv(\comb{id}) & = & \comb{id} &
    \inv(c_1 \seqq c_2) & = & \inv(c_2) \seqq \inv(c_1) \\
    \inv(c_1 + c_2) & = & \inv(c_1) + \inv(c_2) &
    \inv(c_1 \times c_2) & = & \inv(c_1) \times \inv(c_2) \\
    \inv(\comb{swap}^+) & = & \comb{swap}^+ &
    \inv(\comb{swap}^\times) & = & \comb{swap}^\times \\
    \inv(\comb{unit}^+) & = & \comb{uniti}^+ &
    \inv(\comb{uniti}^+) & = & \comb{unit}^+ \\
    \inv(\comb{assoc}^+) & = & \comb{associ}^+ &
    \inv(\comb{associ}^+) & = & \comb{assoc}^+ \\
    \inv(\comb{unit}^\times) & = & \comb{uniti}^\times &
    \inv(\comb{uniti}^\times) & = & \comb{unit}^\times \\
    \inv(\comb{assoc}^\times) & = & \comb{associ}^\times &
    \inv(\comb{associ}^\times) & = & \comb{assoc}^\times \\
    \inv(\comb{distrib}) & = & \comb{distribi} &
    \inv(\comb{distribi}) & = & \comb{distrib} \\
    \inv(\comb{distribo}) & = & \comb{distriboi} &
    \inv(\comb{distriboi}) & = & \comb{distribo} \\
    \inv(\comb{phase}_\varphi) & = & \comb{phase}_{\overline{\varphi}} &
    \inv(\comb{hadamard}) & = & \comb{hadamard}
  \end{array}
\end{equation*}
\end{minipage}
}
\caption{The inversion meta-combinator $\inv$ in $(\mathcal{U})\Pi$.}
\label{fig:inv}
\end{figure}
Our presentation of $\Pi$ differs slightly from \cite{jamessabry:infeff}: our syntax does not include an inversion combinator $\comb{inv}~c$; instead we derive it as a metacombinator (in Fig.~\ref{fig:inv}). This avoids some superfluous syntax -- e.g., $\comb{inv} (c_1 + c_2)$ and $(\comb{inv}~c_1) + (\comb{inv}~c_2)$ are equivalent, as are $\comb{inv}(\comb{id})$ and $\comb{id}$ -- but results in a higher number of base combinators. Some basic well-behavedness properties can be straightforwardly shown by induction, summarised as follows.

\begin{proposition}
  Let $c$ be a $(\mathcal{U})\Pi$ combinator. Then:
  \begin{enumerate}[(i)]
    \item $c \colon b_1 \fromto b_2$ implies $\inv(c) \colon b_2 \fromto b_1$, and
    \item $\inv(\inv(c)) = c$.
  \end{enumerate}
\end{proposition}

\subsection{Reversible quantum combinators: $\UPi$} 
\label{sub:reversible_quantum_combinators_upi}
\begin{figure}
  \fbox{
  \begin{minipage}{0.95\textwidth}
  \textbf{Syntax}
  \begin{align*}
    a & ::= \cdots \mid \comb{phase}_\varphi \mid \comb{hadamard} & 
    \text{(primitive combinators)} \\
    d & ::= \cdots \mid \comb{superposition} & 
    \text{(derived combinators)}
  \end{align*}
  \textbf{Typing rules}
  \begin{equation*}
    \begin{array}{c c c c c c}
      \comb{phase}_\varphi & \of & 1 \fromto 1 & \of & 
      \comb{phase}_{\overline{\varphi}} \\
      \comb{hadamard} & \of & 1+1 \fromto 1+1 & \of & \comb{hadamard} \\
      \comb{superposition} & \of & b+b \fromto b+b & \of & \comb{superposition}
    \end{array}
  \end{equation*}
  \end{minipage}
  }
  \caption{The syntax and typing rules of $\UPi$ in addition to those 
  in $\Pi$ (see Fig.~\ref{fig:pi}).}
  \label{fig:upi}
\end{figure}
$\UPi$ (``yuppie'') extends $\Pi$ with notions of phase and superposition, in the form of the $\comb{phase}_\varphi$ and $\comb{hadamard}$ combinators. Fig.~\ref{fig:upi} shows the syntax and types of this small extension. While the types of $\UPi$ remain the same as in $\Pi$, in $\UPi$ they are entirely quantum rather than (as in $\Pi$) entirely classical. For example, where $1+1$ in $\Pi$ is the type of \emph{bits}, in $\UPi$ it is the type of \emph{qubits} (with no way of forming the type of bits). $\UPi$ canonically takes semantics in the category $\cat{Unitary}$ of finite dimensional Hilbert spaces and unitaries. The full treatment of these semantics is given in Section~\ref{sec:semantics}, but later in this section we will show that $\UPi$ is \emph{approximately universal} for  unitaries.

Phases correspond with unitaries $\mathbb{C} \to \mathbb{C}$. Since $\mathbb{C}$ is the tensor unit in $\cat{Unitary}$, we can express an arbitrary phase $\varphi$ through the combinator $\comb{phase}_\varphi \colon 1 \fromto 1$. This will allow us to form quantum \emph{phase gates} like $S$ and $T$, and to multiply a combinator $c \colon b_1 \fromto b_2$ by an arbitary phase $\varphi$ as:
\[
  \varphi \bullet c = \comb{uniti}^\times \seqq c \times \comb{phase}_\varphi 
  \seqq \comb{unit}^\times
\]
We include \emph{all} phases, yielding an uncountable number of phase
combinators, even though a finite number of phases suffice for
approximate universality up to a global phase. We find including all of them to be the more principled solution, especially when an appropriate finite subset (such as $\{\pm i, \pm 1, \pm \cos(\frac{\pi}{4})
\pm i \sin(\frac{\pi}{4})\}$) can be chosen in a concrete implementation without detriment.

Superpositions are introduced by means of the $\comb{hadamard}$ combinator (of type $1+1 \fromto 1+1$, or $\Qbit \fromto \Qbit$), named after the Hadamard gate from which it takes its semantics:
\[
  \ket0  \mapsto \tfrac{1}{\sqrt{2}}(\ket0 + \ket1) 
  \qquad\qquad
  \ket1  \mapsto \tfrac{1}{\sqrt{2}}(\ket0 - \ket1)
\]
It introduces uniform superpositions of states in the computational basis. Though effective, it can be argued that this combinator is not conceptually clean: all of the $\Pi$ combinators are parametrically polymorphic and pertain to \emph{structure} rather than \emph{behaviour}, but $\comb{hadamard}$ is monomorphic, and pertains specifically to the behaviour of qubits. To mend this, we could instead introduce a parametrically polymorphic combinator $\comb{superposition} : b + b \fromto b + b$ with semantics:
\[
  \inl(\ket \psi) \mapsto \tfrac{1}{\sqrt{2}}(\inl(\ket \psi) +
  \inr(\ket \psi)) 
  \qquad\qquad
  \inr (\ket \psi) \mapsto \tfrac{1}{\sqrt{2}}(\inl(\ket \psi) -
  \inr(\ket \psi)) 
\]
Now $\comb{hadamard}$ is just the $\comb{superposition}$ combinator on the type $1+1 \fromto 1+1$. Interestingly, the two are equivalent in the presence of the other combinators, as $\comb{superposition}$ can be derived:
\[
  \comb{superposition} = (\comb{uniti}^\times + \comb{uniti}^\times) \seqq
  \comb{distribi} \seqq (\comb{id} \times \comb{hadamard}) \seqq \comb{distrib}
  \seqq (\comb{unit}^\times + \comb{unit}^\times) 
\]
It also follows from this definition that $\comb{superposition}$, like $\comb{hadamard}$, is self-inverse. Whether the $\comb{hadamard}$ or the $\comb{superposition}$ combinator is taken as primal thus comes down to preference; there is no difference in expressivity, and one is easily derived from the other.

\subsubsection{Expressiveness}
We have taken an established combinator calculus for reversible \emph{classical} computing, and extended it only slightly with two quantum combinators modelling phase and superposition. One may ask whether this extension is sufficient to express all of reversible \emph{quantum} computing. This question contains a number of subtleties, not least because there are systems of quantum computing which do include concepts of both phase and superposition, but can nevertheless be efficiently simulated by purely classical means (e.g., the Clifford gate set \emph{without} $T$).

Fig.~\ref{fig:gates} shows the implementation of a variety of reversible quantum (and classical) gates in $\UPi$. Note the meta-combinator $\comb{ctrl}$, which produces a combinator for the usual \emph{controlled gate} for a combinator corresponding to a gate $c$. Briefly, $\comb{ctrl}~c$ takes $(\ket0, \ket\psi)$ to $(\ket0, \ket\psi)$, and $(\ket1, \ket\psi)$ to $(\ket1, c(\ket\psi))$. Though $\UPi$ cannot distinguish at the type level between classical and quantum data, it is computationally universal for reversible classical computing, with $\{\comb{px},\comb{cnot},\comb{toffoli}\}$ as one example of a universal gate set~\cite{toffoli:reversible}.
\begin{figure}
  \fbox{
  \begin{minipage}{0.95\textwidth}
  \small
  \begin{tabular}{rcl rcl rcl}
    $\comb{px}$ & $\of$ & $\Qbit \fromto \Qbit$ & 
    $\comb{py}$ & $\of$ & $\Qbit \fromto \Qbit$ &
    $\comb{pz}$ & $\of$ & $\Qbit \fromto \Qbit$ \\
    $\comb{px}$ & $=$ & $\comb{swap}^+$ &
    $\comb{py}$ & $=$ & $\comb{swap}^+ \seqq (\comb{phase}_{-i} + 
    \comb{phase}_i)$ &
    $\comb{pz}$ & $=$ & $\comb{id} + \comb{phase}_{-1}$ \\ \\
    $\comb{s}$ & $\of$ & $\Qbit \fromto \Qbit$ & & & & $\comb{t}$ & $\of$ & 
    $\Qbit \fromto \Qbit$ \\
    $\comb{s}$ & $=$ & $\comb{id} + \comb{phase}_i$ & & & &
    $\comb{t}$ & $=$ & $\comb{id} + \comb{phase}_{e^{\frac{i \pi}{4}}}$ \\ \\
    
    $\comb{ctrl}~c$ & $\of$ & \multicolumn{7}{l}{$b
    \fromto b \to \Qbit \times b \fromto \Qbit \times b$} \\
    $\comb{ctrl}~c$ & $=$ & 
    \multicolumn{7}{l}{$\comb{swap}^\times \seqq \comb{distrib} \seqq (\comb{unit}^\times +
    \comb{unit}^\times) \seqq (\comb{id} + c) \seqq (\comb{uniti}^\times +
    \comb{uniti}^\times) \seqq \comb{distribi} \seqq \comb{swap}^\times$} \\ \\
    
    $\comb{cnot}$ & $\of$ & $\Qbit^2 \fromto \Qbit^2$ &
    $\comb{toffoli}$ & $\of$ & $\Qbit^3 \fromto \Qbit^3$ &
    $\comb{fredkin}$ & $\of$ & $\Qbit^3 \fromto \Qbit^3$ \\
    $\comb{cnot}$ & $=$ & $\comb{ctrl}~\comb{px}$ &
    $\comb{toffoli}$ & $=$ & $\comb{ctrl}~\comb{cnot}$ &
    $\comb{fredkin}$ & $=$ & $\comb{ctrl}~\comb{swap}^\times$
  \end{tabular}
  \normalsize
  \end{minipage}
  }
  \caption{The implementation of a variety of quantum gates in $\UPi$. We use 
  $\Qbit^n$ as shorthand for the $n$-fold product of the qubit type 
  $\Qbit = 1+1$ with itself.}
  \label{fig:gates}
\end{figure}

Returning to the quantum case, using these representations of quantum gates, it can be shown that $\UPi$ is \emph{approximately universal} for reversible quantum computing as well: it can approximate any unitary (on a space of dimension $2^n$) to arbitrary precision (measured by the operator norm). 

\begin{theorem}\label{thm:exp_upi}
  $\UPi$ is approximately universal for $2^n \times 2^n$ unitaries: For any
  unitary $U$ and $\delta > 0$ there exists a $\UPi$ combinator $u$ such that
  $\opnorm{U - \sem{u}} < \delta$.
\end{theorem}

\subsection{Quantum combinators with allocation: the arrow metalanguage $\UPia$}
Next we extend $\UPi$ with an \emph{allocation effect} $\comb{alloc} \of 0 \ato a$, yielding the language of $\UPia$ (\emph{``yuppie-a''}). This effect is introduced by letting combinators $b_1 \ato b_2$ in $\UPia$ be given by invertible combinators with a \emph{heap} of type $b_3$: that is, as $\UPi$ combinators of type $b_1 + b_3 \fromto b_2$. Analogous to \cite{jamessabry:infeff}, this enables the type system to track the information effects.

This small extension will allow us to define a \emph{classical cloning} combinator that clones classical states exactly, and sends quantum states $\ket\psi$ to $\sqrt{\ket{\psi}} \otimes \sqrt{\ket{\psi}}$; this will be crucial later on in deriving a combinator for measurement in $\UPichia$.

$\UPia$ canonically takes semantics in the category $\Isometry$ of Hilbert spaces and isometries: in Section~\ref{sec:semantics}, we will see how a categorical model of $\UPi$ can be extended universally to model of $\UPia$, and Section~\ref{sec:properties} shows how this construction connects the canonical model of $\UPi$ to that of $\UPia$. Now we show that the approximate universality theorem for $\UPi$ with its unitary semantics extends to an approximate universality for $\UPia$ with its semantics in isometries.

\begin{figure}
  \fbox{
  \begin{minipage}{0.95\textwidth}
  \textbf{Syntax}
  \begin{align*}
    b & ::= 0 \mid 1 \mid b + b \mid b \times b & \text{(base types)} \\
    t & ::= b \ato b & \text{(combinator types)} \\
    c & ::= \comb{lift}~u & \text{(primitive combinators)} \\
    d & ::= \comb{iso} \mid \comb{arr}~u \mid c \acmp c \mid
    \comb{first}~c \mid \comb{second}~c \mid \comb{left}~c \mid \comb{right}~c
    \\ & \enspace \mid c \ttt c \mid c \ppp c \mid \comb{inhab} \mid \comb{inl} 
    \mid \comb{inr} \mid \comb{alloc} \mid \comb{clone} & \text{(derived 
    combinators)}
  \end{align*}
  \textbf{Typing rules}
  \begin{equation*}
    \frac{u \of b_1 + b_3 \fromto b_2}{\comb{lift}~u \of b_1 \ato b_2} \quad
    \frac{u \of b_1 \fromto b_2}{\comb{arr}~u \of b_1 \ato b_2} \quad
    \frac{c_1 \of b_1 \ato b_2 \quad c_2 \of b_2 \ato b_3}{c_1 \acmp c_2 
    \of b_1 \ato b_3}
  \end{equation*}
  
  \begin{equation*}
    \frac{c \of b_1 \ato b_2}{\comb{first}~c \of b_1 \times b_3 \ato b_2 \times 
    b_3} \quad 
    \frac{c \of b_1 \ato b_2}{\comb{second}~c \of b_3 \times b_1 \ato b_3
    \times b_2}
  \end{equation*}
  
  \begin{equation*}
    \frac{c \of b_1 \ato b_2}{\comb{left}~c \of b_1 + b_3 \ato b_2 + b_3} \quad 
    \frac{c \of b_1 \ato b_2}{\comb{right}~c \of b_3 + b_1 \ato b_3 + b_2} \quad
  \end{equation*}
  
  \begin{equation*}
    \frac{c_1 \of b_1 \ato b_3 \quad c_2 : b_2 \ato b_4}{c_1 \ppp c_2 \of b_1 + 
    b_2 \ato b_3 + b_4} \quad 
    \frac{c_1 \of b_1 \ato b_3 \quad c_2 : b_2 \ato b_4}{c_1 \ttt c_2 \of b_1
    \times b_2 \ato b_3 \times b_4} \quad
    \frac{b~\text{inhabited}}{\comb{inhab} \of 1 \ato b}
  \end{equation*}
  \begin{equation*}
    \frac{}{\comb{alloc} \of 0 \ato a} \quad
    \frac{}{\comb{inl}   \of a \ato a + b} \quad
    \frac{}{\comb{inr}   \of b \ato a + b} \quad
    \frac{}{\comb{clone} \of a \ato a \times a}
  \end{equation*}
  
  \begin{equation*}
    \frac{}{1~\text{inhabited}} \quad
    \frac{b_1~\text{inhabited} \quad b_2~\text{inhabited}}{b_1 \times 
    b_2~\text{inhabited}} \quad
    \frac{b_1~\text{inhabited}}{b_1 + b_2~\text{inhabited}} \quad
    \frac{b_2~\text{inhabited}}{b_1 + b_2~\text{inhabited}}
  \end{equation*}
  \end{minipage}
  }
  \caption{The syntax and type system of the arrow metalanguage $\UPia$.}
  \label{fig:upia}
\end{figure}
Fig.~\ref{fig:upia} gives an over view of $\UPia$. It is an arrow metalanguage~\cite{hughes:arrows,powerrobinson:premonoidal,jamessabry:infeff} built atop $\UPi$: it has the same base types as $\UPi$, but introduces a new, irreversible combinator type $b \ato b$ (reflecting the fact that combinators in $\UPia$ are no longer invertible). All combinators in $\UPia$ are constructed from combinators in $\UPi$ by means of the single primitive $\comb{lift}$ combinator, following the type rule:
\[
\frac{u \of b_1 + b_3 \fromto b_2}{\comb{lift}~u \of b_1 \ato b_2}
\]
So a combinator in $\UPia$ corresponds to a combinator in $\UPi$ with a hidden \emph{heap} of type $b_3$. Section~\ref{sec:semantics} will discuss that some quotienting is needed for this construction to behave; we defer further details about the semantics until then, including the arrow laws.

For this to constitute an arrow, we must produce meta-combinators $\comb{arr}$, $\acmp$, and $\comb{first}$. To start, $\comb{arr}$ must lift a $\UPi$ combinator $u$ to a \emph{pure} $\UPia$ one, free of effects. To do this, we assign it the trivial heap $0$ and remove it before proceeding with $u$:
\[
  \comb{arr}~u = \comb{lift}(\comb{unit}^+ \seqq u)
\]
In Fig.~\ref{fig:upia}, $\comb{iso}$ refers to atomic combinators of $\UPi$ brought into $\UPia$ by applying $\comb{arr}$ to them. We write, for example, $\comb{swap}^+$ in $\UPia$ to refer to $\comb{arr}(\comb{swap}^+)$, and so on.

To compose combinators $\comb{lift}(u_1) \of b_1 \ato b_2$ and $\comb{lift}(u_2) \of b_2 \ato b_3$ with heaps of type $b_4$ and $b_4'$, we must track both heaps. The result will be a lifted $\UPi$ combinator of type $b_1 + (b_4 + b_4') \fromto b_3$ which permit $u_1$ and $u_2$ access to their parts of the heap accordingly:
\[
  \comb{lift}(u_1) \acmp \comb{lift}(u_2) = \comb{lift}(\comb{associ}^+ \seqq (u_1 + \comb{id}) \seqq u_2)
\]
The final combinator related to $\comb{lift}$ defining an arrow is $\comb{first}$, allowing two arrows (each with their own information effects) to be executed in parallel. We define
$$\comb{lift}(u_1) \ttt \comb{lift}(u_2) = \comb{lift}(\comb{associ}^+ \seqq
\comb{distribi} + \comb{distribi} \seqq \comb{swap}^\times + \comb{swap}^\times \seqq \comb{distribi} \seqq \comb{swap}^\times \seqq u_1 \times u_2)$$
and derive $\comb{first}~c = c \ttt \comb{arr}(\comb{id})$ and $\comb{second}~c = \comb{arr}(\comb{id}) \ttt c$ as usual.
That is, given $\comb{lift}(u_1)$ and $\comb{lift}(u_2)$ with $u_1 \of b_1 + b_3 \fromto b_2$ and $u_2 \of b_1' + b_3' \fromto b_2'$, this defines their product by choosing the heap to be $(b_3 \times b_1') + ((b_1 \times b_3') + (b_3 \times b_3'))$, as we then have:
\begin{align*}
  (b_1 \times b_1') + ((b_1 \times b_3') + ((b_3 \times b_1') + (b_3 \times 
  b_3'))) & \cong
  ((b_1 \times b_1') + (b_1 \times b_3')) + ((b_3 \times b_1') + (b_3 \times 
  b_3')) \\ & \cong
  (b_1 \times (b_1' + b_3')) + (b_3 \times (b_1' + b_3')) \\
  & \cong (b_1 + b_3) \times (b_1' + b_3') 
\end{align*}
Make $\comb{lift}$ an \emph{arrow with choice} by defining a combinator $c_1 \ppp c_2$ giving a choice between $c_1$ and $c_2$:
\[
  \comb{lift}(u_1) \ppp \comb{lift}(u_2) = \comb{lift}((u_1 + u_2) \seqq \comb{midswap}^+)
\]
From this we can derive $\comb{left}~c = c \ppp \comb{arr}(\comb{id})$ and $\comb{right}~c = \comb{arr}(\comb{id}) \ppp c$. 

What can we do with this arrow metalanguage? Firstly, we can construct the promised allocation combinator $0 \ato a$ by lifting the $\UPi$ map $0 + a \fromto a$ that removes the additive unit on the left, i.e., $\comb{alloc} = \comb{lift}(\comb{swap}^+ \seqq \comb{unit}^+)$. From this we can recover injections $\comb{inl}$ and $\comb{inr}$ as
$$\comb{inl} = \comb{arr}(\comb{uniti}^+) \acmp \comb{right}(\comb{alloc})$$ and analogously for $\comb{inr}$, though we can also define them equivalently as $\comb{inl} = \comb{lift}(\comb{id})$ and $\comb{inr} = \comb{lift}(\comb{swap}^+)$. Another crucial application of allocation is \emph{classical cloning}.

\subsubsection{Classical cloning} 
\label{ssub:classical_cloning}
Classical data can be copied, quantum data cannot. While there is a program that inputs a piece of classical data and outputs two copies of that data, no such program exists for quantum data; this is the \emph{no cloning theorem}~\cite{nielsenchuang:quantum,heunenvicary:cqt}.

In light of this, it may come as a bit of a surprise that we can derive a combinator $\comb{clone}$ satisfying $\sem{\comb{clone}}(\ket{0}) = \ket{00}$ and $\sem{\comb{clone}}(\ket{1}) = \ket{11}$. After all, wouldn't that imply $\sem{\comb{clone}}(\ket{\phi}) = \ket{\phi} \otimes \ket{\phi}$ for a qubit $\ket{\phi}$? Fortunately not! To see this, consider a superposed state $\ket{\phi} = \alpha \ket{0} + \beta \ket{1}$. 
\begin{align*}
\sem{\comb{clone}}(\ket{\phi}) & = \sem{\comb{clone}}(\alpha \ket{0} + \beta \ket{1}) = 
\sem{\comb{clone}}(\alpha \ket{0}) + \sem{\comb{clone}}(\beta \ket{1}) \\ & = \alpha\,\sem{\comb{clone}}(\ket{0}) + 
\beta\,\sem{\comb{clone}}(\ket{1}) = \alpha \ket{00} + \beta \ket{11}
\end{align*}
Now, $\ket{00}$ and $\ket{11}$ are shorthands for $\ket{0} \otimes \ket{0}$ and $\ket{1} \otimes \ket{1}$, and the tensor product of Hilbert spaces satisfies $s(\ket{u} \otimes \ket{v}) = (s\ket{u}) \otimes \ket{v} = \ket{u} \otimes (s\ket{v})$ for all scalars $s$ and vectors $\ket{u} \otimes \ket{v}$ in $U \otimes V$. This means for example that $\alpha \ket{00} = (\sqrt{\alpha} \ket{0}) \otimes (\sqrt{\alpha} \ket{0})$, which in general is distinct from $(\alpha \ket{0}) \otimes (\alpha \ket{0})$. So applying $\comb{clone}$ to $\ket{\phi}$ does \emph{not} gives two copies $\ket{\phi} \otimes \ket{\phi}$, but rather two copies of $\ket{\phi}$ with all amplitudes (in the computational basis) replaced by their \emph{square roots}. 

Define $\comb{clone} \of b \ato b \times b$ by induction on the structure of $b$. The base cases are $\comb{clone}_0 = \comb{arr}(\comb{distriboi})$ and $\comb{clone}_1 = \comb{arr}(\comb{uniti}^\times)$. Products are cloned inductively by rearranging:
$$\comb{clone}_{b \times b'} = (\comb{clone}_b \ttt \comb{clone}_b') \acmp \comb{arr}(\comb{midswap}^\times)$$
Sums are cloned inductively, tagging accordingly, and factoring: 
$$\comb{clone}_{b + b'} = (\comb{clone}_b \ppp \comb{clone}_{b'}) \acmp ((\comb{inl} \ppp \comb{id}) \ttt (\comb{inr} \ppp \comb{id})) \acmp \comb{arr}(\comb{distribi}).$$
Interestingly, though the languages and semantics are different, cloning is defined \emph{precisely} as for classical information effects~\cite{jamessabry:infeff}.

\subsubsection{Inhabitation} 
\label{ssub:inhabitation}
Later, we will need a notion of \emph{inhabited types} in $\UPia$. By a type $b$ being inhabited in $\UPia$, we mean that there is a combinator $1 \ato b$. For inhabited types $b$, construct canonical inhabitants as follows. First, $\comb{inhab}_1 = \comb{id}$. The inhabitant of a product type $b \times b'$ is the product $\comb{inhab}_{b \times b'} = \comb{uniti}^\times \acmp \comb{inhab}_b \ttt \comb{inhab}_{b'}$ of inhabitants. Finally, a sum $b + b'$ is inhabited if either $b$ or $b'$ is: if $b$ is inhabited set $\comb{inhab}_{b + b'} = \comb{inhab}_b \acmp \comb{inl}$, and if $b$ is not inhabited but $b'$ is, $\comb{inhab}_{b + b'} = \comb{inhab}_{b'} \acmp \comb{inr}$.

\subsubsection{Expressiveness}
We can now extend the expressiveness theorem for $\UPi$ to one for $\UPia$. 

\begin{theorem}\label{thm:exp_upia}
  $\UPia$ is approximately universal for isometries: For any $2^n \times 2^m$ 
  isometry $V$ and $\delta > 0$ there exists a $\UPia$ combinator $v$ such that
  $\opnorm{V - \sem{v}} < \delta$.
\end{theorem}

\subsection{Quantum combinators with hiding and allocation: the arrow metalanguage $\UPichia$}

\begin{figure}
  \fbox{
  \begin{minipage}{0.95\textwidth}
  \textbf{Syntax}
  \begin{align*}
    b & ::= 0 \mid 1 \mid b + b \mid b \times b & \text{(base types)} \\
    t & ::= b \acto b & \text{(combinator types)} \\
    c & ::= \comb{lift}~v & \text{(primitive combinators)} \\
    d & ::= \comb{iso} \mid \comb{arr}~v \mid c \acmp c \mid
    \comb{first}~c \mid \comb{second}~c \mid \comb{left}~c \mid \comb{right}~c
    \\ 
    & \enspace \mid c \ttt c \mid c \ppp c \mid \comb{inhab} \mid \comb{inl}
    \mid \comb{inr} \mid \comb{alloc} \mid \comb{clone} \\ 
    & \enspace \mid \comb{discard} \mid \comb{fst} \mid \comb{snd} \mid
    \comb{merge} \mid \comb{measure} & \text{(derived combinators)}
  \end{align*}
  \textbf{Typing rules}
  \begin{equation*}
    \frac{v \of b_1 \ato b_2 \times b_3 \quad
    b_3~\text{inhabited}}{\comb{lift}~v \of b_1 \acto b_2} \quad
    \frac{v \of b_1 \ato b_2}{\comb{arr}~v \of b_1 \acto b_2} \quad
    \frac{c_1 \of b_1 \acto b_2 \quad c_2 \of b_2 \acto b_3}{c_1 \acmp c_2 
    \of b_1 \acto b_3}
  \end{equation*}
  
  \begin{equation*}
    \frac{c \of b_1 \acto b_2}{\comb{first}~c \of b_1 \times b_3 \acto b_2
    \times b_3} \quad
    \frac{c \of b_1 \acto b_2}{\comb{second}~c \of b_3 \times b_1 \acto b_3
    \times b_2}
  \end{equation*}
  
  \begin{equation*}
    \frac{c \of b_1 \acto b_2}{\comb{left}~c \of b_1 + b_3 \acto b_2 + b_3} 
    \quad 
    \frac{c \of b_1 \acto b_2}{\comb{right}~c \of b_3 + b_1 \acto b_3 + b_2}   \end{equation*}
  
  \begin{equation*}
    \frac{c_1 \of b_1 \acto b_3 \quad c_2 \of b_2 \acto b_4}{c_1 \ppp c_2 \of
    b_1 + b_2 \acto b_3 + b_4} \quad
    \frac{c_1 \of b_1 \acto b_3 \quad c_2 \of b_2
    \acto b_4}{c_1 \ttt c_2 \of b_1 \times b_2 \acto b_3 \times b_4} \quad
    \frac{b~\text{inhabited}}{\comb{inhab} \of 1 \acto b}
  \end{equation*}
  \begin{equation*}
    \frac{}{\comb{alloc} \of 0 \acto a} \quad
    \frac{}{\comb{inl}   \of a \acto a + b} \quad
    \frac{}{\comb{inr}   \of b \acto a + b} \quad
    \frac{}{\comb{clone} \of a \acto a \times a}
  \end{equation*}
  \begin{equation*}
    \frac{}{\comb{discard} \of a \acto 1} \quad
    \frac{}{\comb{fst}     \of a \times b \acto a} \quad
    \frac{}{\comb{snd}     \of a \times b \acto b} \quad
    \frac{}{\comb{merge}   \of a + a \acto a} \quad
    \frac{}{\comb{measure} \of a \acto a} \quad
  \end{equation*}
  \end{minipage}
  }
  \caption{The syntax and type system of the arrow metalanguage $\UPichia$ 
  (rules for inhabitation appear in Fig.~\ref{fig:upia}).}
  \label{fig:upichia}
\end{figure}
We finally extend $\UPia$ with an additional information effect to \emph{hide} information via a combinator $\comb{discard} \of b \acto 1$, giving us the language of $\UPichia$ (\emph{``yuppie-chi-a''}). Dually to how allocation was introduced in $\UPia$, discarding is introduced in $\UPichia$ by letting combinators $b_1 \acto b_2$ be given by $\UPia$ combinators of type $b_1 \ato b_2 \times b_3$, where we think of $b_3$ as the type of \emph{garbage} produced by the combinator. In order to be able to produce a choice metacombinator, however, we need to make the additional assumption that this garbage is inhabited. This is a very mild assumption, since garbage can always be chosen to be inhabited.

The hiding combinator allows projections $\comb{fst} \of b_1 \times b_2 \acto b_1$ and $\comb{snd} \of b_1 \times b_2 \acto b_2$ to be defined. When combined with the classical cloning combinator inherited from $\UPia$, we show that a combinator $\comb{measure} \of b \acto b$ for measurement can be derived.

$\UPichia$ takes its canonical semantics in the category $\CPTP$ of Hilbert spaces and quantum channels, and as with $\UPia$, we will show in Section~\ref{sec:semantics} how a model of $\UPia$ can be extended to one of $\UPichia$ by a universal construction, connecting isometries to quantum channels (more on this in Section~\ref{sec:properties}). We also extend the approximate universality theorem of $\UPia$ to one showing approximate universality of $\UPichia$ combinators with respect to quantum channels.

Like $\UPia$, $\UPichia$ is an arrow metalanguage extending $\UPia$ (see Fig.~\ref{fig:upichia} for an overview). On the surface, $\UPichia$ may look similar to if we were to apply the arrow construction from the classical case~\cite{jamessabry:infeff} to $\UPi$, but the result would be quite different. The arrow constructions of $\UPia$ and $\UPichia$ are chosen precisely for their semantic properties (which we explore in Section~\ref{sec:semantics}), and cannot be replaced without altering semantics. One tangible difference is in the treatment of allocation: $\mathrm{ML}_\Pi$ (of \cite{jamessabry:infeff}) does not have a unit $0$ for the sum (as that would lead to an undesirable combinator of type $1 \acto 0$), so allocation has type $1 \acto b$; on the other hand, $\UPia$ and $\UPichia$ have a unit $0$ for the sum, and allocation has type $0 \acto b$. 

Similarly to $\UPia$, $\UPichia$ uses the same base types as $\UPi$ and $\UPia$, but introduces a new combinator type $b \acto b$ to distinguish $\UPichia$ combinators at the type level. 
All combinators in $\UPichia$ are constructed from $\UPia$ combinators using the single primitive $\comb{lift}$ combinator
$$
\frac{v \of b_1 \ato b_2 \times b_3 \quad b_3~\text{inhabited}}{\comb{lift}~v \of b_1 \acto b_2} \enspace.
$$
The definition of the arrow metacombinators $\comb{arr}$, $\acmp$, and $\comb{first}$ are bound to look very familiar, as they are defined dually to those in $\UPia$ (indeed, we will see in Section~\ref{sec:semantics} that the two constructions are dual in a formal sense). To turn a $\UPia$ combinator $b_1 \ato b_2$ into a pure $\UPichia$ combinator $b_1 \acto b_2$ can be done by assigning it the trivial (and trivially inhabited) garbage of $1$,
$$
\comb{arr}(v) = \comb{lift}(v \acmp \comb{uniti}^\times) \enspace.
$$
Combinators of type $b_1 \acto b_2$ and $b_2 \acto b_3$ with garbage of type $b_4$ and $b_4'$ respectively can be composed by
$$
\comb{lift}(v_1) \acmp \comb{lift}(v_2) = \comb{lift}(v_1 \acmp (v_2 \ttt \comb{id}) \acmp \comb{assoc}^\times)
$$
resulting in a combinator with garbage $b_4' \times b_4$. For the final arrow combinator $\comb{first}$ allowing parallel execution of arrows, we define $c_1 \ttt c_2$ to simply run the underlying $\UPia$ combinators in parallel and swap the garbage into the right position as necessary,
$$\comb{lift}(v_1) \ttt \comb{lift}(v_2) = \comb{lift}((v_1 \ttt v_2) \acmp \comb{midswap}^\times) \enspace,$$
such that the garbage of $c_1 \ttt c_2$ is the product of the garbages of $c_1$ and $c_2$ respectively. We derive $\comb{first}(c) = c \ttt \comb{id}$ and $\comb{second}(c) = \comb{id} \ttt c$. All of these definitions are straightforwardly seen to preserve the inhabitation requirement on garbage.

Defining the choice metacombinator $c_1 \ppp c_2$ is a bit more tricky, and it turns out to be easier to define $\comb{left}$ and derive $\ppp$ and $\comb{right}$ from it. The idea is to exploit distributivity and inhabitation of garbage: if $c \of b_1 \acto b_2$ produces garbage of type $b_4$ and the identity produces garbage of type $1$, we can use the inhabitation of $b_4$ to turn the trivial garbage into garbage of type $b_4$ via $\comb{inhab} \of 1 \acto b_4$, and then distribute out on the right to get something of the required type $(b_2 + b_3) \times b_4$. This gives us the definition
$$ \comb{left}(\comb{lift}~v) = \comb{lift}((v \ppp \comb{uniti}^\times) \acmp (\comb{id} \ppp (\comb{id} \ttt \comb{inhab})) \acmp (\comb{swap}^\times \ppp \comb{swap}^\times) \acmp \comb{distribi} \acmp \comb{swap}^\times)$$
from which we derive $\comb{right}$ and $\ppp$ as usual~\cite{hughes:arrows} as $$\comb{right}(c) = \comb{swap}^+ \acmp \comb{left}(c)
\acmp \comb{swap}^+ \qquad c_1 \ppp c_2 = \comb{left}(c_1) \acmp \comb{right}(c_2) \enspace.$$
The combinators $\comb{alloc}$, $\comb{inl}$, $\comb{inr}$, $\comb{clone}$, and $\comb{inhab}$ related to the allocation effect from $\UPia$, as well as all of the base combinators of $\UPi$ lifted to $\UPia$, can be further lifted to $\UPichia$ by applying $\comb{arr}$ to them (these are denoted by $\comb{iso}$ in Fig.~\ref{fig:upichia}).

Information hiding is introduced in $\UPichia$ by means of the effectful $\comb{discard} \of b \acto 1$ combinator. Some finesse is required to manage the inhabitation requirement on garbage, however. On all types aside from $0$,  $\comb{discard}$ is given by lifting the $\UPia$ combinator $b \ato 1 \times b$ that adds the multiplicative unit on the left,
$$ \comb{discard} = \comb{lift}(\comb{uniti}^\times \acmp \comb{swap}^\times) \enspace.$$
On $0$, we first need to use $\comb{alloc}$ to allocate something of inhabited type, namely $1$, before we can discard it:
$$ \comb{discard} = \comb{lift}(\comb{alloc} \acmp \comb{uniti}^\times) \enspace.$$
Analogously to the injections in $\UPia$, we can derive \emph{projections} from this discarding effect as
$$ \comb{fst} = \comb{id} \ttt \comb{discard} \acmp \comb{unit}^\times \qquad
\comb{snd} = \comb{swap}^\times \acmp \comb{fst}$$
though these can also be defined equivalently as $\comb{fst} = \comb{lift}(\comb{id})$ and $\comb{snd} = \comb{lift}(\comb{swap}^\times)$.

To allow the choice metacombinator to be used for conditional execution, we need a way to merge branches. This can be defined in $\UPichia$ as the $\comb{merge} \of b + b \acto b$ combinator, exploiting hiding and the fact that $a + a \cong a \times (1+1)$, as in
$$ \comb{merge} = (\comb{uniti}^\times \ppp \comb{uniti}^\times) \acmp \comb{distribi} \acmp \comb{fst} \enspace.$$
We are finally ready to explore the measurement combinator in $\UPichia$.

\subsubsection{Measurement in the computational basis} 
\label{ssub:measurement}
As we have seen previously, $\UPia$ permits a notion of classical cloning, and $\UPichia$ inherits it. When we combine this with the ability to discard information in $\UPichia$ using the projections, we obtain a surprisingly robust notion of measurement in the computational basis. This measurement combinator $\comb{measure} \of b \acto b$ is defined simply to be
$$
  \comb{measure} = \comb{clone} \acmp \comb{fst}
$$
Classically, this is a complicated way of doing absolutely nothing -- the map takes a piece of classical data, copies it, and then immediately throws away the copy. In the quantum case, however, this performs measurement. We illustrate this by an example.

Consider an arbitrary qubit state $\ket{\phi} = \alpha \ket{0} + \beta \ket{1}$, and the associated density matrix $$\dyad{\phi}{\phi} = (\alpha \ket{0} + \beta \ket{1})(\overline{\alpha} \bra{0} + \overline{\beta} \bra{1})$$
Conjugating by $\sem{\comb{clone}}$ (noting that this does indeed perform classical cloning in the canonical model of $\cat{CPTP}$) yields the density matrix
\begin{align}
 \sem{\comb{clone}} \dyad{\phi}{\phi} \sem{\comb{clone}}^\dagger & = (\alpha \ket{00} + \beta \ket{11})(\overline{\alpha} \bra{00} + \overline{\beta} \bra{11}) \notag\\
 & \label{eq:meas_clone} = |\alpha|^2 \dyad{00}{00} + \alpha\overline{\beta} \dyad{00}{11} + \beta\overline{\alpha} \dyad{11}{00} + |\beta|^2 \dyad{11}{11} \enspace.
\end{align}
We remark that, in $\cat{CPTP}$, $\sem{\comb{fst}}$ is given by the \emph{partial trace} (see, e.g., \cite{nielsenchuang:quantum}) of density matrices. This means that
\begin{align}
& \sem{\comb{fst}}(|\alpha|^2 \dyad{00}{00} + \alpha\overline{\beta} \dyad{00}{11} + \beta\overline{\alpha} \dyad{11}{00} + |\beta|^2 \dyad{11}{11}) \notag\\ & \qquad = 
|\alpha|^2 \tr(\dyad{0}{0})\dyad{0}{0} + \alpha\overline{\beta} \tr(\dyad{0}{1})\dyad{0}{1} + \beta\overline{\alpha} \tr(\dyad{1}{0})\dyad{1}{0} + |\beta|^2 \tr(\dyad{1}{1})\dyad{1}{1}
\notag\\ & \label{eq:meas_trace}\qquad = |\alpha|^2 \dyad{0}{0} + |\beta|^2 \dyad{1}{1}
\end{align}
since for vectors $\ket{a}$ and $\ket{b}$ in the computational basis,  $\tr(\dyad{a}{b}) = 1$ when $\ket{a} = \ket{b}$, and $\tr(\dyad{a}{b}) = 0$ otherwise. Note that measurement in an arbitrary basis can then be performed by conjugating the measurement combinator with the appropriate change-of-base combinator. For example, measurement in the Hadamard basis $\{\ket+,\ket-\}$ is performed using $\comb{hadamard} \acmp \comb{measure} \acmp \comb{hadamard}$.

This method of measurement may seem counterintuitive, but it is important to note that \eqref{eq:meas_clone} and \eqref{eq:meas_trace} above show that it is physically equivalent to the usual one. Our presentation can be seen as exploiting \emph{purification} of quantum states (see, e.g., \cite{nielsenchuang:quantum}) to describe mixed states, as this result implies that every mixed state appears as the partial trace of a pure one.

\subsubsection{Expressiveness}
Finally we can extend the universality theorems for $\UPi$ and $\UPia$ to one that $\UPichia$ is approximately universal for arbitrary quantum computations, that is, quantum channels. 

\begin{theorem}\label{thm:exp_upichia}
  $\UPichia$ is approximately universal for quantum channels: For any quantum 
  channel $\Lambda$ and $\delta > 0$ there exists a $\UPichia$ combinator $c$ 
  such that $\opnorm{\Lambda - \sem{c}} < \delta$.
\end{theorem}

\section{Categorical semantics}\label{sec:semantics}

In this section we develop denotational semantics for the simple programming languages of the previous section in three stages. First, the base language $\Pi$ can be interpreted in rig groupoids. This is then extended to $\UPi$ by providing interpretations for the phase and Hadamard combinators. Finally, we discuss two categorical constructions, $R$ and $L$, that model the arrow constructions -- i.e., $\comb{lift}$ combinator constructors of $\UPia$ and $\UPichia$ respectively -- in order to encapsulate information allocation and hiding and account for measurement. 

\subsection{Rig groupoids}

To interpret the type system and semantics of $\Pi$ in a category, it needs to have combinators $\oplus$ and $\otimes$ that distribute over each other in the correct way. This is captured in the notion of a \emph{rig groupoid}.
Recall that a groupoid is a category in which every morphism is invertible.

\begin{definition}
  A \emph{rig category} is category $\cat{C}$ with two symmetric monoidal structures $(\oplus,0)$ and $(\otimes,I)$, as well as natural isomorphisms
  \begin{align*}
    \delta_{A,B,C} \colon A \otimes (B \oplus C) &\to (A \otimes B) \oplus (A \otimes C) \\
    \delta_0 \colon A \otimes 0 &\to 0 
  \end{align*}
  satisfying several coherence laws for which we refer to~\cite{laplaza:coherence,kelly:coherence}.
\end{definition}

Here $\oplus$ need not be a coproduct, and $\otimes$ need not be a product, which is the special case of a \emph{distributive category}.

A rig groupoid suffices to interpret $\Pi$. Being a language for classical reversible computing, a canonical such model is the rig groupoid $\FinBij$ of finite sets and bijective functions.
The base types are interpreted as $\sem{0} = 0$, $\sem{1}=I$, $\sem{b+b'}=\sem{b} \oplus \sem{b'}$, and $\sem{b \times b'} = \sem{b} \otimes \sem{b'}$.
The combinator type $b \fromto b'$ becomes the (invertible) morphisms $\sem{b} \to \sem{b'}$.
As for the atomic combinators, the swap morphisms for $\oplus$ and $\otimes$ interpret $\comb{swap}^+$ and $\comb{swap}^\times$, respectively. 
The unitor $\rho^\oplus_A \colon A \oplus \to 0$ and its inverses coming from the monoidal structure $(\oplus,0)$ denote $\comb{unit}^+$ and $\comb{uniti}^+$.
The associator $\alpha^\oplus_{A,B,C} \colon (A \oplus B) \oplus C \to A \oplus (B \oplus C)$ and its inverse interpret $\comb{assoc}^+$ and $\comb{associ}^+$.
We use the coherence isomorphisms of $(\otimes,I)$ to interpret $\sem{\comb{unit}^\times} = \rho^\otimes$, $\sem{\comb{uniti}^\times} = (\rho^\otimes)^{-1}$, $\sem{\comb{assoc}^\times}=\alpha^\otimes$, and $\sem{\comb{associ}^\times}=(\alpha^\otimes)^{-1}$.
The distributors finish the interpretation: $\sem{\comb{distrib}} = \delta$, $\sem{\comb{distribi}} = \delta^{-1}$, $\sem{\comb{distribo}}= \delta_0$, and $\sem{\comb{distriboi}}= \delta_0^{-1}$.
The combinators are simply composition via $\circ$, $\oplus$, and $\otimes$. 

Interpreting $\UPi$ requires additionally to give semantics to the $\comb{phase}_\varphi$ and $\comb{hadamard}$ combinators. In principle, they can be given trivial semantics (i.e., as identities) in any rig groupoid, though that would yield mere classical semantics, and thus defeat the purpose of the quantum extension to $\Pi$ to begin with. A far better approach would be to interpret $\UPi$ in a category with quantum capabilities, the canonical choice being the category $\Unitary$ of finite dimensional Hilbert spaces and unitaries. Here, the semantics of these two combinators are given by
\begin{align*}
  \sem{phase_\varphi} \quad & = \quad x \mapsto \varphi \cdot x &
  \sem{hadamard} \quad & = \quad \frac{1}{\sqrt{2}} \left( \begin{matrix}
    1 & 1 \\
    1 & -1
  \end{matrix} \right) \enspace.
\end{align*}
Another way of writing $\sem{phase_\varphi}$ is as the $1 \times 1$ matrix $(\varphi)$, as phases are exactly unitaries $\mathbb{C} \to \mathbb{C}$.

\subsection{Garbage and heap}

We now extend the categorical semantics of $\UPi$ to take quantum information effects into account.
Let $\cat{C}$ be a symmetric monoidal category. We think of its objects as types, its morphisms as programs, and its tensor product as parallel composition. Hence a morphism $f \colon A \to B \otimes G$ will denote a program that takes an input of type $A$ and produces an output of type $B$ together with some \emph{garbage} of type $G$. As we want to disregard the garbage, we will identify this morphism with $(\id_B \otimes h) \circ f \colon A \to B \otimes G'$ for any morphism $h \colon G \to G'$ that postprocesses the garbage, called a \emph{mediator}. That is, we consider the equivalence relation $\sim_L$ generated by:
\begin{equation}\label{eq:simL}
  \begin{tikzpicture}
	\begin{pgfonlayer}{nodelayer}
		\node [style=none] (13) at (3.25, 0) {$\sim_L$};
		\node [style=map] (14) at (1, -0.25) {$f$};
		\node [style=none] (15) at (0, 2.5) {};
		\node [style=none] (16) at (0, 0) {};
		\node [style=none] (17) at (2, 0) {};
		\node [style=none] (18) at (0, -0.5) {};
		\node [style=none] (19) at (0, -1.5) {};
		\node [style=none] (22) at (2, 2.5) {};
		\node [style=none] (23) at (-0.25, -1.25) {\tiny $A$};
		\node [style=none] (25) at (-0.25, 2.25) {\tiny $B$};
		\node [style=none] (26) at (1.75, 2.25) {\tiny $G$};
		\node [style=map] (28) at (5.75, -0.25) {$f$};
		\node [style=none] (29) at (4.75, 2.5) {};
		\node [style=none] (30) at (4.75, 0) {};
		\node [style=none] (31) at (6.75, 0) {};
		\node [style=none] (32) at (4.75, -0.5) {};
		\node [style=none] (33) at (4.75, -1.5) {};
		\node [style=none] (36) at (6.75, 1.25) {};
		\node [style=none] (37) at (4.5, -1.25) {\tiny $A$};
		\node [style=none] (39) at (4.5, 2.25) {\tiny $B$};
		\node [style=none] (40) at (6.5, 0.75) {\tiny $G$};
		\node [style=none] (55) at (6.75, 1.75) {};
		\node [style=none] (56) at (6.75, 2.5) {};
		\node [style=none] (57) at (6.5, 2.25) {\tiny $G'$};
		\node [style=map1] (58) at (6.75, 1.5) {$h$};
	\end{pgfonlayer}
	\begin{pgfonlayer}{edgelayer}
		\draw (15.center) to (16.center);
		\draw (18.center) to (19.center);
		\draw (22.center) to (17.center);
		\draw (29.center) to (30.center);
		\draw (32.center) to (33.center);
		\draw (36.center) to (31.center);
		\draw (56.center) to (55.center);
	\end{pgfonlayer}
\end{tikzpicture}
\end{equation}

Dually, instead of garbage, we can also consider a \emph{heap}. That is, morphisms $f \colon A \oplus H \to B$ will denote a program that takes input of type $A$ and may use a heap $H$ in producing an output of type $B$. Again, we only care about access to the heap and not the actual contents of the heap, so we will identify $f$ with $f \circ (\id_A \oplus h) \colon A \oplus H' \to B$ for any morphisms $h \colon H' \to H$ that preprocesses the heap. That is, we consider the equivalence relation $\sim_R$ generated by:
\begin{equation}\label{eq:simR}
  \begin{tikzpicture}
	\begin{pgfonlayer}{nodelayer}
		\node [style=none] (13) at (3.25, 1.25) {$\sim_R$};
		\node [style=map] (14) at (1, 1.25) {$f$};
		\node [style=none] (15) at (0, -1.5) {};
		\node [style=none] (16) at (0, 1) {};
		\node [style=none] (17) at (2, 1) {};
		\node [style=none] (18) at (0, 1.5) {};
		\node [style=none] (19) at (0, 2.75) {};
		\node [style=none] (22) at (2, -1.5) {};
		\node [style=none] (23) at (-0.25, 2.5) {\tiny $B$};
		\node [style=none] (25) at (-0.25, -1.25) {\tiny $A$};
		\node [style=none] (26) at (1.75, -1.25) {\tiny $H$};
		\node [style=map] (28) at (5.75, 1.25) {$f$};
		\node [style=none] (29) at (4.75, -1.5) {};
		\node [style=none] (30) at (4.75, 1) {};
		\node [style=none] (31) at (6.75, 1) {};
		\node [style=none] (32) at (4.75, 1.5) {};
		\node [style=none] (33) at (4.75, 2.75) {};
		\node [style=none] (36) at (6.75, -0.25) {};
		\node [style=none] (37) at (4.5, 2.5) {\tiny $B$};
		\node [style=none] (39) at (4.5, -1.25) {\tiny $A$};
		\node [style=none] (40) at (6.5, 0.25) {\tiny $H$};
		\node [style=none] (55) at (6.75, -0.75) {};
		\node [style=none] (56) at (6.75, -1.5) {};
		\node [style=none] (57) at (6.5, -1.25) {\tiny $H'$};
		\node [style=map1] (58) at (6.75, -0.5) {$h$};
	\end{pgfonlayer}
	\begin{pgfonlayer}{edgelayer}
		\draw (15.center) to (16.center);
		\draw (18.center) to (19.center);
		\draw (22.center) to (17.center);
		\draw (29.center) to (30.center);
		\draw (32.center) to (33.center);
		\draw (36.center) to (31.center);
		\draw (56.center) to (55.center);
	\end{pgfonlayer}
\end{tikzpicture}
\end{equation}
Notice that the diagram in \eqref{eq:simL} refers to the monoidal product $\otimes$, while the one above in \eqref{eq:simR} refers to the monoidal sum $\oplus$.

What exactly are $\sim_L$ and $\sim_R$? Equations~\eqref{eq:simL} and~\eqref{eq:simR} define relations that are already reflexive (with the identity as mediator) and transitive (compose mediators), but not always symmetric. The following lemma shows that they are already symmetric in special cases of interest, such as when the base category $\cat{C}$ is a groupoid.

\begin{lemma}
  When every morphism in $\cat{C}$ is split monic, equation~\eqref{eq:simL} defines an equivalence relation.
  When every morphism in $\cat{C}$ is split epic, equation~\eqref{eq:simR} defines an equivalence relation.
\end{lemma}
\begin{proof}
  It suffices to establish symmetry. If $(\id \otimes h) \circ f = g$, there is $k$ with $k \circ h = \id$, so $(\id \otimes k) \circ g = (\id \otimes k) \circ (\id \otimes h) \circ f = f$. An analogous argument holds for $\sim_R$.
\end{proof}

\begin{proposition}\label{prop:L}
  If $(\cat{C},\otimes,I)$ is a symmetric monoidal category, there is a well-defined symmetric monoidal category $\LL{\cat{C}}$ whose:
  \begin{itemize}
    \item \emph{objects} are the same as those of $\cat{C}$;
    \item \emph{morphisms} $A \to B$ are equivalence classes of morphisms $A \to B \otimes G$ in $\cat{C}$ under~\eqref{eq:simL};
    \item \emph{composition} is: \ctikzfig{compl}
    \item \emph{identities} are the inverse right unitors: \ctikzfig{idl}
    \item \emph{tensor unit} $I$ is as in $\cat{C}$;
    \item \emph{tensor product of objects} is as in $\cat{C}$;
    \item \emph{tensor product of morphisms} is: \ctikzfig{tensorl}
  \end{itemize}
  Dually, if $(\cat{C},\oplus,O)$ is a symmetric monoidal category, there is a well-defined symmetric monoidal category $\RR{\cat{C}} = \LL{\cat{C}\opp}\opp$. Explicitly:
  \begin{itemize}
    \item \emph{objects} are the same as those of $\cat{C}$;
    \item \emph{morphisms} $A \to B$ are equivalence classes of morphisms $A \oplus H \to B$ in $\cat{C}$ under~\eqref{eq:simR}.
  \qed
  \end{itemize}
\end{proposition}
\begin{proof}
  Well-definedness follows from Lemma~\ref{lem:Rwelldef}.
  It is straightforward to verify that coherence isomorphisms in $\LL{\cat{C}}$  may be taken to be those in $\cat{C}$ composed with the inverse  right unitor.
\end{proof}

\begin{lemma}\label{lem:Rwelldef}
  Let $(\cat{C},\oplus)$ be a monoidal category. If $f \sim_R f'$ and $g \sim_R g'$, then:
  \begin{enumerate}[(i)]
    \item $g \circ f \sim_R g' \circ f'$ if $g$ and $f$ are composable in 
    $\RR{\cat{C}}$;
    \item $f \oplus g \sim_R f' \oplus g'$;
  \end{enumerate}
  Dually $g \circ f \sim_L g' \circ f'$ when $g$ and $f$ composable in 
  $\LL{\cat{C}}$ and $f \otimes g \sim_L f' \otimes g'$ when $f \sim_L f'$ and $g \sim_L g'$ in a monoidal category $(\cat{C},\otimes)$.
\end{lemma}

So what is the point of doing these constructions? This is made clear in the following proposition. Indeed, in Section~\ref{sec:properties}, we will see that $\LL{\cat{C}}$ and $\RR{\cat{C}}$ are, in a certain sense, the smallest categories containing $\cat{C}$ with these properties.

\begin{proposition}
  The monoidal unit $I$ is terminal in $\LL{\cat{C}}$, and the monoidal unit 
  $O$ is initial in $\RR{\cat{C}}$. 
\end{proposition}

As a consequence of this fact, there are canonical projections $X \otimes Y \tot{\pi_1} X$ and $Y \otimes Y \tot{\pi_2} X$ in $\LL{\cat{C}}$ given by $X \otimes Y \tot{\id \otimes {!}} X \otimes I \tot{\rho^\otimes} X$  (and symmetrically for $\pi_2$). Likewise, there are canonical injections $X \tot{\amalg_1} X \oplus Y$ and $Y \tot{\amalg_2} X \oplus Y$ (defined dually to the above) in $\RR{\cat{C}}$.

\begin{proposition}\label{prop:DE}
  If $(\cat{C},\otimes)$ is a monoidal category, there is a strict monoidal functor $\EE \colon \cat{C} \to \LL{\cat{C}}$ given by $\EE(A)=A$ on objects, and on morphisms as:
  \[
    \EE(A \stackrel{f}{\to} B)
    = A \stackrel{f}{\to} B \stackrel{{\rho^{\otimes}}^{-1}}{\to} B \otimes I
  \]
  Dually, if $(\cat{C},\oplus)$ is a monoidal category, there is a strict monoidal functor $\DD \colon \cat{C} \to \RR{\cat{C}}$ given by $\DD(A)=A$ on objects, and on morphisms as:
  \[
   \DD(A \stackrel{f}{\to} B) 
    = A \oplus 0 \stackrel{\rho^{\oplus}}{\to} A \stackrel{f}{\to} B \\
  \]
\end{proposition}
\begin{proof}
  By definition $\EE(A \otimes B) = \EE(A) \otimes \EE(B)$ on objects. On morphisms:
  \ctikzfig{monfunlc}
  Coherence isomorphisms in $\LL{\cat{C}}$ are precisely the image under $\EE$ of those in $\cat{C}$.
\end{proof}

\subsection{Lifting tensor products}
We will compose the $L$ and $R$ constructions, applying them to the base rig groupoid consecutively. In this section we show that if $\cat{C}$ is a rig category then so is $\RR{\cat{C}}$, while $\LL{\RR{\cat{C}}}$ loses its direct sum and becomes merely a monoidal category. However, we will see later that the direct sum in $\LL{\RR{\cat{C}}}$ is \emph{binoidal} and satisfies the laws associated with an arrow with choice.

The overall idea with these constructions is that if $\cat{C}$ interprets the base language $\UPi$, then $\RR{\cat{C}}$ interprets the arrow metalanguage $\UPia$ of $\UPi$ extended with allocation. More precisely, if $\sem{u} \colon \sem{b_1} \oplus \sem{b_3} \to \sem{b_2}$ is the interpretation of some $\UPi$ combinator $u$ in a rig groupoid $\cat{C}$, the interpretation $\sem{\comb{lift}(u)}$ in $\UPia$ is given by the equivalence class $\left[\sem{u}\right]_{\sim_R} \colon \sem{b_1} \to \sem{b_2}$ in $\RR{\cat{C}}$. In turn, $\LL{\RR{\cat{C}}}$ interprets the arrow metalanguage $\UPichia$ extending $\UPi$ with allocation and hiding by interpreting $\sem{\comb{lift}(v)} \colon \sem{b_1} \to \sem{b_2}$ as the equivalence class of $\sem{v} \colon \sem{b_1} \to \sem{b_2} \otimes \sem{b_3}$ in $\RR{\cat{C}}$.
Later, in Section~\ref{subsec:universalproperties} we will exhibit universal properties of $\RR{\cat{C}}$ and $\LL{\cat{C}}$, and argue that they justify these constructions in the canonical semantics of $\UPi$: $\RR{\Unitary}$ is equivalent to the category $\Isometry$ of finite dimensional Hilbert spaces and isometries, while $\LL{\RR{\Unitary}}$ is equivalent to the category $\CPTP$ of finite dimensional Hilbert spaces and quantum channels.

Although $L$ and $R$ are dual constructions, the order in which they transform $\cat{C}$ is important: we first add heaps and then garbage. The reason for this asymmetry is the following lemma.

\begin{lemma}\label{lem:Linitialobject}
  Let $\cat{C}$ be a symmetric monoidal category. If $\cat{C}$ has an initial object, then so does $\LL{\cat{C}}$.
\end{lemma}
\begin{proof}
  Let $0$ be an initial object in $\cat{C}$. 
  For each object $A$ there is then a morphism $0 \to A$ in $\LL{\cat{C}}$ given by the morphism 
  \ctikzfig{init_lc} 
  in $\cat{C}$. If $f \colon 0 \to A$ is a morphism in $\LL{\cat{C}}$, represented by a morphism $f \colon 0 \to A \otimes G$ in $\cat{C}$, then:
  \ctikzfig{init_lc2} 
  Thus $0$ is indeed initial in $\LL{\cat{C}}$ as well.
\end{proof}

We start by endowing $\otimes$ and $\oplus$ with the capability to allow heaps.

\begin{lemma}\label{lem:Rrig}
  If $\cat{C}$ is a rig category, then so is $\RR{\cat{C}}$.
\end{lemma}
\begin{proof}[Proof sketch]
  The monoidal structure $(\oplus,0)$ is inherited from $\cat{C}$ straightforwardly. More intricately, $\RR{\cat{C}}$ is monoidal with $(\otimes,I)$ inherited from $\cat{C}$:
  \begin{itemize}
    \item the tensor product of objects is $A \otimes B$;
    \item the tensor unit is $I$;
    \item the tensor product of morphisms $f \colon A \oplus H \to B$ and $f' \colon A' \oplus H' \to B'$ is
    \[
      (A \otimes A') \oplus H'' \stackrel{\delta^{-1}}{\to} (A \oplus H) \otimes (A' \oplus H') \stackrel{f \otimes f'}{\longrightarrow} B \otimes B'
    \]
    where $H'' = (H \otimes A') \oplus (A \otimes H') \oplus (H \otimes H')$;
  \end{itemize}
\end{proof}

The dual result does not hold: if $\cat{C}$ is a rig category, then $\LL{\cat{C}}$ need not be (though it is always monoidal by Proposition~\ref{prop:L}). This asymmetry is caused by the fact that $\otimes$ distributes over $\oplus$, but not the other way around. Lemma~\ref{lem:Linitialobject} is the nullary case of this fact.
For a special case where this does hold, regard a Boolean algebra as a posetal category $\cat{C}$; then $\wedge$ and $\vee$ do distribute over each other both ways, so in that case $\LL{\cat{C}}$ is again a rig category. 

\begin{lemma}\label{lem:DDrig}
  If $\cat{C}$ is a rig category, then $\DD \colon \cat{C} \to \RR{\cat{C}}$ is a strict rig functor.
\end{lemma}
\begin{proof} 
  By definition $\DD(A \otimes B) = \DD(A) \otimes \DD(B)$ on objects. On morphisms, $\DD(f) \otimes \DD(g)$ is
  \[
    (A \otimes A') \oplus H'' \tot{\delta} (A \oplus 0) \otimes (A' \oplus 0)
    \tot{\rho^\oplus \otimes \rho^\oplus} A \otimes A'
    \tot{f \otimes g} B \otimes B'
  \]
  with $H'' = (0 \otimes A') \oplus (A \otimes 0) \oplus (0 \otimes 0)$. Now $H'' \simeq 0$ by successive applications of the annihilators and unitors for $\oplus$, so by coherence $\DD(f) \otimes \DD(g)$ is the composition of $\rho^\oplus \colon (A \otimes A') \oplus 0 \to A \otimes A'$ and $f \otimes g$, which is precisely $\DD(f \otimes g)$.
\end{proof}

\subsection{Arrows with choice}\label{subsec:arrows}

To complete the categorical semantics, we will show that $\LL{\RR{\cat{C}}}$ supports \emph{arrows with choice}~\cite{hughes:arrows}. We saw in Proposition~\ref{prop:L} and Lemma~\ref{lem:Rrig} that if $\cat{C}$ is a rig category, then $\LL{\RR{\cat{C}}}$ is monoidal under $\otimes$. What happens to $\oplus$? It is no longer necessarily a monoidal structure on $\LL{\RR{\cat{C}}}$, but we will show that it is \emph{binoidal} -- roughly, it is monoidal except that $\oplus$ is only a functor in each variable separately rather than in both variables jointly~\cite{powerrobinson:premonoidal} (see also \cite{jacobsheunenhasuo:arrows}). 

\begin{definition}
  A category $\cat{C}$ is \emph{binoidal} when it is equipped with  functors $A \oplus - \colon \cat{C} \to \cat{C}$ and $ - \oplus B \colon \cat{C} \to \cat{C}$ for each choice of objects $A,B \in \cat{C}$ such that applying the first functor to $B$ results in the same object $A \oplus B$ as applying the second functor to $A$.

  An \emph{arrow with choice} is a functor $F$ from a monoidal category $(\cat{C},\oplus,0)$ where $0$ is initial to a binoidal category $(\cat{D},\oplus)$ such that for any $g \colon B \to B'$:
  \begin{align}
    F(f \oplus B) &= F(f) \oplus B \label{eq:arrowchoice1}\\
    F(\amalg_1) \circ f &= (f \oplus B) \circ F(\amalg_1) \label{eq:arrowchoice2}\\
    (f \oplus B') \circ F(A \oplus g) &= F(A \oplus g) \circ (f \oplus B) \label{eq:arrowchoice3}\\
    \alpha_\oplus \circ ((f \oplus B) \oplus C) &= (f \oplus (B \oplus C)) \circ \alpha_\oplus \label{eq:arrowchoice4}
  \end{align} 
\end{definition}

To prove that $\LL{\RR{\cat{C}}}$ is binoidal under $\oplus$, we will require that it has \emph{inhabited garbage}: for any equivalence class of morphisms $A \to B \otimes G$ in $\RR{\cat{C}}$, there is a morphism $I \to G$ in $\RR{\cat{C}}$. Because $0$ is initial in $\RR{\cat{C}}$, the problematic case where there is an isomorphism $\vartheta \colon G \to 0$ is handily avoided as  $f \colon A \to B \otimes G$ is then equivalent to $A \tot{f} B \otimes G \tot{\id \otimes \vartheta} B \otimes 0 \tot{\id \otimes {!}} B
\otimes I$. 

In the remaining cases, inhabited garbage can be constructed when $\cat{C}$ is \emph{semisimple}, meaning that any object is isomorphic to one built out of copies of the tensor unit $I$ using $\oplus$ and $\otimes$. This is the case for $\Unitary$. In this case, canonical inhabitants, i.e., morphisms $I \to G$ for each inhabited object $G$, can be constructed as the interpretation of the $\UPia$ combinators given in Section~\ref{ssub:inhabitation}.

\begin{lemma}\label{lem:binoidal}
  If $\cat{C}$ is a semisimple rig category, then $\LL{\RR{\cat{C}}}$ is a binoidal category under $\oplus$.
\end{lemma}

\begin{proposition}\label{prop:arrowchoice}
  If $\cat{C}$ is a semisimple rig category, $\EE \colon \RR{\cat{C}} \to \LL{\RR{\cat{C}}}$ is an arrow with choice.
\end{proposition}

Combining the previous Proposition with Lemma~\ref{lem:DDrig} shows that $\EE \circ \DD \colon \cat{C} \to \LL{\RR{\cat{C}}}$ is an arrow with choice. The categorical semantics of $\UPi$, $\UPia$, and $\UPichia$, including certain derived combinators with structural or otherwise significant interpretations, are summarised in Fig.~\ref{fig:semantics}.
 
\begin{figure}
  \fbox{
  \begin{minipage}{0.95\textwidth}
  \textbf{Base types}\hfill\phantom{x}\small
  \begin{align*}
    \begin{array}{llll}
      \sem{0} = O & \sem{1} = I & \sem{b+b'} = \sem{b} \oplus \sem{b'} & 
      \sem{b \times b'} = \sem{b} \otimes \sem{b'}
    \end{array}
  \end{align*}
  \normalsize
  
  \textbf{Semantics of $\UPi$} \hfill
  \small\colorbox{palegreen}{$\sem{b_1 \fromto b_2} = \sem{b_1} \to \sem{b_2} 
  \text{ in \cat{C}}$}
  
  \begin{align*}
    & \begin{array}{lllll}
      \sem{\comb{swap}^+} = \sigma^\oplus & 
      \sem{\comb{unit}^+} = \rho^\oplus &
      \sem{\comb{uniti}^+} = (\rho^\oplus)^{-1} &
      \sem{\comb{assoc}^+} = \alpha^\oplus &
      \sem{\comb{associ}^+} = (\alpha^\oplus)^{-1} \\
      \sem{\comb{swap}^\times} = \sigma^\otimes &
      \sem{\comb{unit}^\times} = \rho^\otimes &
      \sem{\comb{uniti}^+} = (\rho^\otimes)^{-1} &
      \sem{\comb{assoc}^\times} = \alpha^\otimes &
      \sem{\comb{associ}^\times} = (\alpha^\otimes)^{-1} \\
      \sem{\comb{id}} = \id &
      \sem{\comb{distrib}} = \delta & 
      \sem{\comb{distribi}} = \delta^{-1} &
      \sem{\comb{distribo}} = \delta_0 &
      \sem{\comb{distriboi}} = \delta_0^{-1}
    \end{array} \\
    & \begin{array}{llll}
      \sem{c_1 \seqq c_2} = \sem{c_2} \circ \sem{c_1} &
      \sem{c_1 + c_2} = \sem{c_1} \oplus \sem{c_2} &
      \sem{c_1 \times c_2} = \sem{c_1} \otimes \sem{c_2} &
      \colorbox{paleyellow}{$\sem{\comb{inv}(c)} = \sem{c}^{-1}$}
    \end{array} \\
    & \colorbox{silver}{
      $\begin{array}{ll}
        \text{In \cat{Unitary}:} \\
        \sem{\comb{phase}_\varphi} = (\varphi) & \sem{\comb{hadamard}} = 
        \frac{1}{\sqrt{2}} \begin{pmatrix}
          1 & 1 \\ 1 & -1
        \end{pmatrix}
      \end{array}$
    }
  \end{align*}
  \normalsize
  
  \textbf{Semantics of $\UPia$} \hfill
  \small\colorbox{palegreen}{$\sem{b_1 \ato b_2} = \sem{b_1} \to \sem{b_2} 
  \text{ in $\RR{\cat{C}}$}$}
  
  \begin{align*}
    & \begin{array}{lll}
      \sem{\comb{lift}~u} = \left[\sem{u}\right]_{\sim_R} &
      \colorbox{paleyellow}{$\sem{\comb{arr}~v} = \DD(\sem{v})$} &
      \colorbox{paleyellow}{$\sem{c_1 \acmp c_2} = \sem{c_2} \circ \sem{c_1}$} 
      \\
      \colorbox{paleyellow}{$\sem{c_1 \ttt c_2} = \sem{c_1} \otimes \sem{c_2}$} 
      & 
      \colorbox{paleyellow}{$\sem{c_1 \ppp c_2} = \sem{c_1} \oplus \sem{c_2}$} &
      \colorbox{paleyellow}{$\sem{\comb{first}~c} = \sem{c} \otimes \id$} \\
      \colorbox{paleyellow}{$\sem{\comb{second}~c} = \id \otimes \sem{c}$} & 
      \colorbox{paleyellow}{$\sem{\comb{left}~c} = \sem{c} \oplus \id$} &
      \colorbox{paleyellow}{$\sem{\comb{right}~c} = \id \oplus \sem{c}$} \\
      \colorbox{paleyellow}{$\sem{\comb{alloc}} = O \tot{!} \sem{b_2}$} & 
      \colorbox{paleyellow}{$\sem{\comb{inl}} = \amalg_1$} &
      \colorbox{paleyellow}{$\sem{\comb{inr}} = \amalg_2$} \\
    \end{array}
  \end{align*}
  \normalsize
  
  \textbf{Semantics of $\UPichia$} \hfill
  \small\colorbox{palegreen}{$\sem{b_1 \acto b_2} = \sem{b_1} \to \sem{b_2} 
  \text{ in $\LL{\RR{\cat{C}}}$}$}
  \begin{align*}
    & \begin{array}{lll}
      \sem{\comb{lift}~v} = \left[\sem{v}\right]_{\sim_L} &
      \colorbox{paleyellow}{$\sem{\comb{arr}~v} = \EE(\sem{v})$} &
      \colorbox{paleyellow}{$\sem{c_1 \acmp c_2} = \sem{c_2} \circ \sem{c_1}$} 
      \\
      \colorbox{paleyellow}{$\sem{c_1 \ttt c_2} = \sem{c_1} \otimes \sem{c_2}$} 
      & 
      \colorbox{paleyellow}{$\sem{c_1 \ppp c_2} = \sem{\comb{right}~c_2} \circ 
      \sem{\comb{left}~c_1}$} &
      \colorbox{paleyellow}{$\sem{\comb{first}~c} = \sem{c} \otimes \id$} \\
      \colorbox{paleyellow}{$\sem{\comb{second}~c} = \id \otimes \sem{c}$} & 
      \colorbox{paleyellow}{$\sem{\comb{left}~c} = \sem{c} \oplus 
      \sem{b_1'}$} &
      \colorbox{paleyellow}{$\sem{\comb{right}~c} = \sem{b_1'} \oplus 
      \sem{c}$} \\
      \colorbox{paleyellow}{$\sem{\comb{discard}} = \sem{b_1} \tot{!} I$} & 
      \colorbox{paleyellow}{$\sem{\comb{fst}} = \pi_1$} &
      \colorbox{paleyellow}{$\sem{\comb{snd}} = \pi_2$} \\
    \end{array}
  \end{align*}
  \normalsize
  \end{minipage}}
  \caption{The categorical semantics of $\UPi$, $\UPia$, and $\UPichia$ in
  summary, including the semantics of various derived combinators (marked
  yellow). The semantics of $\comb{phase}_\varphi$ and $\comb{hadamard}$ in
  $\UPi$ refer to their canonical semantics in $\Unitary$. In $\UPichia$,
  $b_1'$ in the semantics of $\comb{left}~c$ and $\comb{right}~c$ refer to the
  type of the alternate choice; e.g., when $c$ has type $b_1 \acto b_2$,
  $\comb{left}~c$ has type $b_1 + b_1' \acto b_2 + b_1'$.}
  \label{fig:semantics}
\end{figure}

\section{Properties of the categorical constructions}\label{sec:properties}

The previous section developed enough properties of our categorical semantics to interpret $\UPia$. But the $L$- and $R$-constructions have more properties, that give them the status of a very useful generic construction. From these universal properties it follows that two fundamental embeddings in reversible computing and quantum computing are both captured by our categorical semantics.

\subsection{Universal properties}\label{subsec:universalproperties}

The first insight is the following factorisation lemma, that brings morphisms in $\LL{\cat{C}}$ and $\RR{\cat{C}}$ in a normal form in terms of pure morphisms from $\cat{C}$.

\begin{lemma}\label{lem:factor}
  If $\cat{C}$ is a rig category, then:
  \begin{enumerate}[(i)]
    \item any map $A \to B$ in $\RR{\cat{C}}$ is represented by 
    $A \tot{\amalg_1} A \oplus H \tot{\DD(f)} B$ for some $f \colon A 
    \oplus H \to B$ in $\cat{C}$;
    \item any map $A \to B$ in $\LL{\cat{C}}$ is represented by $A \tot{\EE(f)} B \otimes G \tot{\pi_1} B$ for some $f \colon A \to B \otimes G$ in $\cat{C}$.
  \end{enumerate}
  These factorisations are unique.
\end{lemma}
\begin{proof}
  Point (ii) follows from~\cite[Lemma~8]{hermidatennent:indeterminates}, see~\cite[Lemma~15]{heunenkaarsgaard:bennettstinespring}.
  Point (i) then follows from (ii) by Proposition~\ref{prop:L}.
\end{proof}

For any object $A$ there is a unique morphism $A \to I$ in $\LL{\cat{C}}$, represented by the unitor $A \to I \otimes A$ in $\cat{C}$, that simply considers the input as garbage and does nothing else. This makes $\LL{\cat{C}}$ affine, and in fact it is the universal affine category including $\cat{C}$.

\begin{theorem}\label{thm:affine}
  $\LL{\cat{C}}$ is the affine completion of a monoidal category $\cat{C}$: for an affine monoidal category $\cat{D}$ and a monoidal functor $F \colon \cat{C} \to \cat{D}$ there is a unique monoidal functor $\hat{F} \colon \LL{\cat{C}} \to \cat{D}$ with $F = \hat{F} \circ \EE$.
\end{theorem}
\begin{proof}
  This is an easy generalisation of~\cite[Theorem~16]{heunenkaarsgaard:bennettstinespring}.
\end{proof}

\begin{theorem}\label{thm:coaffine}
  $\RR{\cat{C}}$ is the coaffine completion of a rig category $\cat{C}$: for any coaffine rig category $\cat{D}$ and rig functor $F \colon \cat{C} \to \cat{D}$ there is a unique rig functor $\hat{F} \colon \RC \to \cat{D}$ such that $F = \hat{F} \circ \DD$.
\end{theorem}
\begin{proof}
  By Proposition~\ref{prop:L} and Theorem~\ref{thm:affine}, there is a unique functor $\hat{F}$ making the triangle commute that is monoidal with respect to $\oplus$. 
  Lemmas~\ref{lem:DDrig} and~\ref{lem:factor} show that it is also monoidal with respect to $\otimes$, that is, a rig functor.
\end{proof}

It follows immediately that the $L$- and $R$-constructions are functorial: a monoidal functor $\cat{C} \to \cat{D}$ induces a monoidal functor $\LL{\cat{C}} \to \LL{\cat{D}}$, and a rig functor $\cat{C} \to \cat{D}$ induces a rig functor $\RR{\cat{C}} \to \RR{\cat{D}}$.
The combination $\LL{\RR{\cat{C}}}$ is more than the sum of Theorems~\ref{thm:affine} and~\ref{thm:coaffine}. The following lemma shows that it is also universal for arrows with choice as in Section~\ref{subsec:arrows}.

\begin{lemma}\label{lem:util}
  Let $\cat{C}$ be a category with an initial object. 
  \begin{enumerate}[(i)]
    \item The category $\LL{\cat{C}}$ has an initial object, and $\EE \colon \cat{C} \to \LL{\cat{C}}$ preserves it.
    \item If $F \colon \cat{C} \to \cat{D}$ preserves the initial object then so does $\hat{F} \colon \LL{\cat{C}} \to \cat{D}$.
    \item If $F \colon \cat{C} \to \cat{D}$ preserves injections then so does $\hat{F} \colon \LL{\cat{C}} \to \cat{D}$.
  \end{enumerate}
\end{lemma}
\begin{proof}
  For (i), observe that any morphism $0 \to A$ in $\LL{\cat{C}}$ must be represented by the unique morphism $0 \to A \otimes G$ in $\cat{C}$ for some $G$, and these are all equivalent under $\sim_L$. By construction $\EE(0)=0$, see Proposition~\ref{prop:DE}.
  Points (ii) and (iii) now follow immediately.
\end{proof}

\subsection{Toffoli and Stinespring}

A central question of foundational importance for reversible modes of computing concerns that of \emph{reversible expressivity}. Any irreversible computing machine is trivially able to simulate a reversible one -- after all, reversible operations are just ordinary operations which happen to be invertible -- but what is lost by considering \emph{only} the reversible ones? Fortunately for the viability of reversible modes of computing, the answer turns out to be ``nothing at all,'' if one is willing to accept some \emph{ancillary inputs} and \emph{garbage outputs} to occur.

\subsubsection{The fundamental theorem of classical reversible computing}

In the case of classical reversible circuit logic, Toffoli found this question to be of so supreme importance to his theory that he dubbed the expressivity theorem the \emph{fundamental theorem}~\cite{toffoli:reversible} of classical reversible computing. Toffoli showed this by demonstrating that any finite function $\phi$ can be simulated by a network consisting of an encoder, a finite bijective function $f$, and a decoder. This encoder and decoder should be \emph{``essentially independent of $\phi$ and contain as little ``computing power'' as possible,''}~\cite{toffoli:reversible} and though this characterisation is (perhaps intentionally) vague, Toffoli goes on to show that one can always choose a ``trivial encoder'' (an injection) and a ``trivial decoder'' (a projection).

It is perhaps surprising that this highly specific statement (and its proof) on the nature of finite functions, down to decomposition of a finite function as an injection, a bijective function, and a projection, can be formulated as the following purely categorical statement:

\begin{theorem} 
  The monoidal functor $\LR{\FinBij} \to \FinSet$ induced by the (monoidal) 
  inclusion functor $\FinBij \to \FinSet$ is full.
\end{theorem}

To clarify, the inclusion functor $\FinBij \tot{I} \FinSet$ acts as the identity on both objects and morphisms, noting that any finite bijection is trivially also a finite function. The monoidal functor $\LR{\FinBij} \to \FinSet$ then arises by successively applying Theorems~\ref{thm:coaffine} and \ref{thm:affine}, as in
\[\begin{tikzcd}
	\FinBij & {\RR{\FinBij}} & {\LR{\FinBij}} \\
	& \FinSet & \FinSet
	\arrow["\id"', from=2-2, to=2-3]
	\arrow["{\hat{\check{I}}}", dotted, from=1-3, to=2-3]
	\arrow["{\check{I}}", dotted, from=1-2, to=2-2]
	\arrow["\EE", from=1-2, to=1-3]
	\arrow["\DD", from=1-1, to=1-2]
	\arrow["I"', from=1-1, to=2-2]
\end{tikzcd}\]

Notice that it follows from Lemma~\ref{lem:factor} that any morphism of
$\LR{\FinBij}$ factors (essentially uniquely) as $A \tot{\EE(\amalg_1)} A
\oplus H \tot{\EE(\DD(f))} B \otimes G \tot{\pi_1} B$ for some bijection $A
\oplus H \tot{f} B \otimes G$. Recall that all morphisms in $\FinBij$ are isomorphisms, and that all functors preserve isomorphisms. Using Lemma~\ref{lem:util} and noting
that $\RR{\FinBij} \tot{\check{I}} \FinSet$ preserves injections, it follows that any function in the image of $\hat{\check{I}}$ factors as
$$A \tot[\mathit{injection}]{\amalg_1} A \oplus H \tot[\mathit{bijection}]{\hat{\check{I}}(\EE(\DD(f)))} B \otimes G \tot[\mathit{projection}]{\pi_1} B$$
What the fundamental theorem of reversible computing then states is that \emph{all} functions are, in fact, in the image of $\hat{\check{I}}$, and so permit such a factorisation. That is exactly to say that $\hat{\check{I}}$ is full.

\subsubsection{Stinespring's dilation theorem}

Similarly, in the setting of quantum theory, the $L$- and $R$-constructions capture Stinespring's dilation theorem. As discussed in Section~\ref{ssub:pure_vs_mixed_quantum_theory}, this theorem shows that any mixed quantum operation can be modeled by a pure quantum operation, if one is willing to enlarge the situation with an auxiliary system.
Write $\Isometry$ for the category of Hilbert spaces and isometric linear maps, and $\Unitary$ for the subcategory of unitary linear maps.

\begin{theorem}[Huot \& Staton]\label{thm:stinespring}
  There is an equivalence $\RR{\Unitary} \simeq \Isometry$ of rig categories, and an equivalence $\LR{\Unitary} \simeq\CPTP$ of monoidal categories.
\end{theorem}
\begin{proof}
  See~\cite{huotstaton:universal,huotstaton:completion} and~\cite{heunenkaarsgaard:bennettstinespring}.
\end{proof}

Notice that the previous theorem holds for Hilbert space that are not necessarily finite-dimensional.

\section{Applications}\label{sec:applications}
We will now consider some areas where our results can be put to work. First, we argue that our categorical model can be used to prove useful, nontrivial properties about measurement entirely algebraically, without ever needing to consider the gritty details of quantum channels. Second, we illustrate the use of $\UPichia$ as a metalanguage by providing a translation from Selinger's quantum flowcharts~\cite{selinger:qpl} to $\UPichia$.

\subsection{Properties of measurement}
In this section, we will use type subscripts on combinators whenever it is necessary to disambiguate between definitions, or to make the presentation clearer. For example, we will write $\comb{measure}_{b+b'}$ to mean the measurement combinator on the type $b+b' \acto b+b'$.

Our first property is bit technical, concerning the behaviour of injections with measurement. We will see shortly how it can be used to prove far more interesting things. 

\begin{proposition}\label{prop:meascomminj}
  Measurement commutes with injections: $\sem{\comb{measure}_b \acmp
  \comb{inl}} = \sem{\comb{inl} \acmp \comb{measure}_{b+b'}}$, and likewise for
  $\comb{inr}$.
\end{proposition}
Any complex, finite dimensional Hilbert space is isomorphic to one of the form $\mathbb{C}^n$, where $n$ is its dimension. From this it follows that each canonical injection $\amalg_i : \mathbb{C} \to \mathbb{C}^n$ is associated with a distinct vector $\ket{i}$ linearly independent from all other $\ket{j}$ for $1 \le j \le n$, $j \neq i$. Together, these are precisely the classical states forming the computational basis of $\mathbb{C}^n$.

Abstracting, we say that a classical state is nothing but an injection, i.e., composition of $\comb{inl}$ and $\comb{inr}$. This intuition is correct, in that we can show that measurement does nothing to classical states:
\begin{proposition}
  If $s$ is a classical state then $\sem{\comb{s} \acmp \comb{measure}} =
  \sem{s}$.
  \begin{proof}
    We first see that measurement on $1$ is the identity: 
    $\comb{measure}_1 = \comb{clone}_1 \acmp \comb{fst} = 
    \comb{uniti}^\times \acmp (\comb{id} \times \comb{discard}) \acmp 
    \comb{unit}^\times$.
    Since $\sem{\comb{discard}}$ when applied to $\sem{1} = I$ is nothing but 
    the identity by $I$ terminal, it follows that 
    \begin{align*}
      \sem{\comb{measure}_1} & = \sem{\comb{uniti}^\times \acmp (\comb{id} 
      \times \comb{discard}) \acmp \comb{unit}^\times} =
      \sem{\comb{unit}^\times} \circ \sem{\comb{id} \times \comb{discard}} \circ
      \sem{\comb{uniti}^\times} \\ & = \sem{\comb{unit}^\times} \circ
      \sem{\comb{uniti}^\times} = \sem{\comb{id}} \enspace.
    \end{align*}
    Now, since a classical state $s \of 1 \acto b$ is precisely an injection, 
    it follows by Proposition~\ref{prop:meascomminj} that 
    $\sem{s \acmp \comb{measure}_b} = \sem{\comb{measure}_1 \acmp s} = \sem{s}
    \circ \sem{\comb{measure_1}} = \sem{s}$.
  \end{proof}
\end{proposition}
A very useful property of measurement is that the result of measuring a joint system is nothing but the product of measurements on each constituent system individually. This is shown as follows:
\begin{proposition}\label{prop:measprod}
  Measurement of products is the product of measurements:
  $\sem{\comb{measure}_{b \times b'}} = \sem{\comb{measure}_b \times
  \comb{measure}_{b'}}$.
\end{proposition}
\begin{proof}
  Using the fact that $\comb{measure} = \comb{clone} \acmp \comb{fst}$ and
  $\comb{clone}_{b \times b'} = \comb{clone}_b \times \comb{clone}_{b'} \acmp
  \comb{midswap}^\times$, the property follows by naturality of 
  $\sem{\comb{midswap}^\times}$:
  \[\begin{tikzcd}[column sep=22mm]
  	{A \otimes B} & {(A \otimes A) \otimes (B \otimes B)} & {(A \otimes I) \otimes (B \otimes I)} \\
  	{(A \otimes A) \otimes (B \otimes B)} & {A \otimes B} \\
  	{(A \otimes B) \otimes (A \otimes B)} && {(A \otimes B) \otimes (I \otimes I)}
  	\arrow["{\sem{\comb{clone}} \otimes \sem{\comb{clone}}}"{pos=0.4}, from=1-1, to=1-2]
  	\arrow["{\sem{\comb{fst}}}", from=3-1, to=2-2]
  	\arrow["{\sem{\comb{fst}}}"', from=3-3, to=2-2]
  	\arrow["{\sem{\comb{id}} \otimes (\sem{\comb{discard}} \otimes \sem{\comb{discard}})}"', from=3-1, to=3-3]
  	\arrow["{\sem{\comb{discard}} \otimes \sem{\comb{discard}} }"{pos=0.6}, from=1-2, to=1-3]
  	\arrow["{\sem{\comb{fst}} \otimes \sem{\comb{fst}}}", from=1-2, to=2-2]
  	\arrow["{\sem{\comb{clone}} \otimes \sem{\comb{clone}}}"', from=1-1, to=2-1]
  	\arrow["{\sem{\comb{midswap}^\times}}"', from=2-1, to=3-1]
  	\arrow["{\sem{\comb{midswap}^\times}}", from=1-3, to=3-3]
  \end{tikzcd}\]
  Again, $A$ and $B$ range over interpretations of arbitrary $\UPichia$ types 
  $b$ and $b'$.
\end{proof}
An immediate consequence of this property is that measurements also commute with projections:
\begin{proposition}
  Measurement on a product type commutes with projections:
  $\sem{\comb{measure_{b \times b'}} \acmp \comb{fst}} = \sem{\comb{fst} \acmp
  \comb{measure}_b}$ and likewise for $\comb{snd}$.
\end{proposition}
\begin{proof}
  Since $\sem{\comb{measure_{b \times b'}}} = \sem{\comb{measure}_b \times
  \comb{measure}_{b'}} = \sem{\comb{measure}_b} \otimes 
  \sem{\comb{measure}_{b'}}$ by Proposition~\ref{prop:measprod}, it follows by 
  naturality of $\pi_1 = \sem{\comb{fst}}$ that
  $$
    \sem{\comb{measure_{b \times b'}} \acmp \comb{fst}} = 
    \pi_1 \circ \sem{\comb{measure}_b} \otimes \sem{\comb{measure}_{b'}} =
    \sem{\comb{measure}_b} \circ \pi_1 = \sem{\comb{fst} \acmp \comb{measure}_b}
    \enspace.
  $$
  which was what we wanted.
\end{proof}
The final property we want to show is that measurement is idempotent. Conceptually, this can be seen as an extension to the property that measurement does nothing to classical states. This is because the result of measuring a quantum state will always be a mixed classical state, so further measuring this has no effect. To do this, we remark that cloning can be shown to be associative.
\begin{proposition}\label{prop:coassoc}
  Cloning is associative: $\sem{\comb{clone} \acmp (\comb{clone} \times 
  \comb{id}) \acmp \comb{assoc}^\times} = \sem{\comb{clone} \acmp (\comb{id} 
  \times \comb{clone})}$.
\end{proposition}
We can then show idempotence:
\begin{proposition}
  Measurement is idempotent: $\sem{\comb{measure} \acmp \comb{measure}} = 
  \sem{\comb{measure}}$.
\end{proposition}
\begin{proof}
  Since $\sem{\comb{fst}} = \sem{\comb{lift}~\comb{id}}$, it can be shown that
  $\sem{\comb{measure}} = \sem{\comb{lift}~\comb{clone}}$, so by the definition
  of $\acmp$ in $\UPichia$ we have that $$\sem{\comb{measure} \acmp
  \comb{measure}} = \sem{\comb{lift}(\comb{clone} \acmp (\comb{clone} \times
  \comb{id}) \acmp \comb{assoc}^\times)} = \sem{\comb{lift}(\comb{clone} \acmp 
  (\comb{id} \times \comb{clone}))},$$
  the final equality following from coassociativity of cloning
  (Proposition~\ref{prop:coassoc}). But then $\sem{\comb{id} \times
  \comb{clone}} = \id \otimes \sem{\comb{clone}}$ mediates between
  $\sem{\comb{clone}}$ and $\sem{\comb{clone} \acmp (\comb{id} \times
  \comb{clone})}$ in $\RR{\cat{C}}$, so $\sem{\comb{measure}} =
  \sem{\comb{lift}~\comb{clone}} = \sem{\comb{lift}~(\comb{clone} \acmp
  (\comb{id} \times \comb{clone}))} = \sem{\comb{measure} \acmp
  \comb{measure}}$ in $\LR{\cat{C}}$.
\end{proof}

\subsection{Quantum flow charts}
\label{sub:qfcs}
In this section, we demonstrate the translation of (noniterative) quantum flow charts into $\UPichia$. As the name suggests, quantum flow charts are the quantum extension of the classical imperative flow chart languages, used extensively in areas such as program compilation and partial evaluation~\cite{jonesgomard:pe,hatcliff:pe}. They were first considered by Selinger in several variations~\cite{selinger:qpl}: here, we consider the purely quantum variant, which has only quantum data, in the form of qubit ensembles. The only type of data supported is the type $\Qbit$ of qubits, so typing contexts $\Gamma$ are simply given by lists of active $\Qbit$ variables. In its textual form, a quantum flowchart is simply a list of commands. The supported commands are as follows (with $q$ ranging over variables):
\begin{align*}
  c \mathbin{::=} \mathit{new~qbit}~q \mathbin{:=} 0 \mid \mathit{discard}~q \mid E \mathrel{*{=}} U \mid \mathit{permute}~\varphi \mid
  \mathit{initial} \mid \mathit{measure}~q \mid \mathit{merge} \mid c; c \mid c \oplus c
\end{align*}
Commands take sums of typing contexts to sums of typing contexts, each summand denoting a program branch. Briefly, $\mathit{new~qbit}$ and $\mathit{discard}$ allocate and discard variables respectively, $E \mathrel{*{=}} U$ applies a unitary $U$ to a non-empty list of variables $E = p,q,r,\dots$, $\mathit{permute}~\varphi$ changes the variable order by applying an arbitrary permutation $\varphi$, $\mathit{initial}$ initialises an empty typing context, and $\mathit{merge}$ merges two program branches (with the same typing context) into one. The most novel command is arguably $\mathit{measure}~q$, which measures the qubit $q$ and branches on the measurement result: this style of measurement-based flow control goes by the motto of ``quantum data, classical control.'' Flow charts are composed in sequence and in parallel using $;$ and $\oplus$ respectively.

We begin with the translation of types and contexts, given simply by $\trans{q : \Qbit} = 1+1$ and $\trans{\Gamma,\Gamma'} = \trans{\Gamma}\times\trans{\Gamma'}$. Before we proceed with the translation of commands, we will use the abuse of notation $\ket0$ and $\ket1$ to refer to the injections $\inl : 1 \acto 1+1$ and $\inr : 1 \acto 1+1$ respectively, to indicate that these serve as allocation of the constant classical values of $\ket0$ and $\ket1$. Commands are translated as follows:
\begin{align*}
\begin{array}{rcl rcl}
  \trans{\mathit{new~qbit}~q \mathrel{:=} 0} \enspace & = & \enspace 
  \comb{uniti}^\times \acmp \comb{id} \times \ket0 &
  \trans{\mathit{discard}~q} \enspace & = & \enspace \comb{fst} \\
  \trans{E \mathrel{*{=}} U} \enspace & = & \enspace \comb{id} \times
  \comb{arr}(\comb{arr}(\hat{U})) &
  \trans{\mathit{permute}~\varphi} \enspace & = & \enspace 
  \comb{arr}(\comb{arr}(\hat{\varphi})) \\
  \trans{\mathit{initial}} \enspace & = & \enspace \comb{alloc} &
  \trans{\mathit{merge}} \enspace & = & \enspace \comb{merge} \\
  \trans{c ; c'} \enspace & = & \enspace \trans{c} \acmp \trans{c'} &
  \trans{c \oplus c'} \enspace & = & \enspace \trans{c} \ppp \trans{c'}
\end{array} \\
\trans{\mathit{measure}~q} \enspace = \enspace \comb{id} \times \comb{measure} \acmp \comb{distrib} \acmp ((\comb{id} \times \ket0) \ppp  (\comb{id} \times \ket1))
\end{align*}
In the above, the terms $\hat{U}$ and $\hat{\varphi}$ denote the approximations of the unitary $U$ and permutation $\varphi$ (a particularly simple kind of unitary) respectively as $\UPi$ terms, as given by Theorem~\ref{thm:exp_upi}.

\section{Conclusion and future work}\label{sec:conclusion}
We have shown how quantum measurement, often presented in somewhat mysterious terms, arises through two suprisingly simple arrow constructions, associated with the information effects of \emph{allocation} and \emph{hiding}, on a reversible quantum combinator language. We have provided categorical semantics for all of these languages through elementary universal constructions on \emph{rig categories}. These let us prove useful nontrivial properties of measurement as semantic equivalences, recast the fundamental theorem of reversible computation as an elementary categorical statement, and interpret (noniterative) quantum flow charts. There are several avenues for further research:

\subsubsection*{Classical and quantum data}
In the current formulation, $\UPi$ only allows forming quantum data types. Hence qubit measurement, for example, can only be given the type $\Qbit \to \Qbit$, rather than $\Qbit \to \Bit$. Extending $\UPichia$ with classical data would address this shortcoming. Semantically, this would need a (sufficiently nice) construction of the category of C*-algebras and quantum channels from the category of Hilbert spaces and quantum channels.

\subsubsection*{Categorical semantics of SILQ} 
The quantum programming language SILQ~\cite{bichselbaadergehrvechev:silq} has several original features: \emph{measurement-free} and \emph{quantum-free} functions, a linear type system, and the automatic uncomputation of garbage.
$\UPichia$ enables a type-level interpretation of \emph{measurement-free} functions (i.e., \emph{pure} combinators), and access to a canonical reversibilisation of combinators. It would be interesting to extend the $\UPi$ family with the remaining features to provide combinator semantics for SILQ, much like $\Pi^0$ does in the classical case for Theseus~\cite{jamessabry:theseus}.

\subsubsection*{Is there a quantum effect?}
We have shown that there are elementary arrow constructions connecting reversible and irreversible quantum computations. Is there a similar arrow construction connecting \emph{classical} reversible computation and \emph{quantum} reversible computation? Such a construction would likely involve several steps, such as adjoining the circle group to introduce phases, and considering a variation of convex combinations of morphisms respecting direct sums, to give morphisms that introduce and eliminate superpositions.

\subsubsection*{Recursion and subnormalised channels}
The notion of quantum channel considered in this paper is too rigid to enable recursion or iteration, which is why we were only able to give semantics to the noniterative fragment of quantum flow charts. To enable recursion requires relaxing the notion of quantum channel, from (completely positive) trace-preserving maps to trace-nonincreasing ones. Can recursion be added as an effect, i.e., an arrow construction from $\cat{CPTP}$ to $\cat{CPTN}$?

\subsubsection*{Measurement-based quantum computation}
Measurement-based quantum computation is in a sense opposite to the quantum circuit model. In the latter all operations are reversible except the very last one. In the former all operations are irreversible up to reversible corrections being fed forward. What is the relationship to the model given in this article? Is there a translation from the language of measurement patterns used in measurement-based quantum computation to $\UPichia$?

\ifarxiv
\vspace{\baselineskip}

\noindent
\textbf{Acknowledgements:} We thank Pablo Andrés-Martínez for his comments and suggestions on this paper. Some early ideas relating to this paper were explored by the second author in conjunction with Martti Karvonen, during a short-term scientific mission to the University of Edinburgh sponsored by COST Action IC1405: \emph{Reversible computation - extending horizons of computing}.
\fi

\bibliography{refs}

\ifarxiv
\appendix

\section{Supplementary material}
\input{supplement_content}
\fi

\end{document}